%% file: main.tex
\documentclass[letterpaper, 10 pt, journal, twoside]{IEEEtran}
\ifCLASSINFOpdf
\else
\fi
\def\BibTeX{{\rm B\kern-.05em{\sc i\kern-.025em b}\kern-.08em
    T\kern-.1667em\lower.7ex\hbox{E}\kern-.125emX}}
\usepackage{balance}
\usepackage{amssymb}
\usepackage{amsmath,amsfonts}
\usepackage{algorithmic}
\usepackage{algorithm}
\usepackage{array}
\usepackage{textcomp}
\usepackage{url}
\usepackage{graphicx}
\usepackage{cite}
\usepackage[caption=false]{subfig}
\newcommand{\argmin}{\mathop{\rm arg~min}\limits}
\usepackage{bm}
\usepackage{here}
\usepackage{amsthm}
\newtheorem{theorem}{Theorem}[section]
\newtheorem{lemma}[theorem]{Lemma}
\newtheorem{remark}{Remark}

\newtheorem{definition}{Definition}

\newtheorem{assumption}{Assumption}

\usepackage{graphicx}
\usepackage{textcomp}
\usepackage{xcolor}
\usepackage{lipsum}% Just for this example
\usepackage{graphicx}
\usepackage{textcomp}
\newlength{\FigWidth}
\usepackage{comment}
\begin{document}
%\onecolumn
%\input{List_of_modifications_IEEE_RA_L_NODA_March_26th_2026/text_RevisionRemark_NODA.tex}
%\twocolumn
\title{Neural Power-Optimal Magnetorquer Solution for\\
Multi-Agent Formation and Attitude Control}
%\begin{comment}
%名前と所属は、原稿、付録、補足資料のどこにも含めないでください。
%人々や資金提供機関への謝辞は、原稿が受理された後にのみ記載してください。
%\end{comment}
% Use only for final RAL version.
%\markboth{Journal of \LaTeX\ Class Files,~Vol.~18, No.~9, September~2020}%
%{How to Use the IEEEtran \LaTeX \ Templates}
% The paper headers
%\markboth{IEEE ROBOTICS AND AUTOMATION LETTERS, Journal of \LaTeX\ Class Files,~Vol.~14, No.~8, August~2021
%}%
%{Shell \MakeLowercase{\textit{et al.}}: A Sample Article Using IEEEtran.cls for IEEE Journals}
%\IEEEpubid{0000--0000/00\$00.00~\copyright~2021 IEEE}
% Remember, if you use this you must call \IEEEpubidadjcol in the second
% column for its text to clear the IEEEpubid mark.
\author{Yuta Takahashi$^{1}$, Shin-ichiro Sakai$^{2}$
\thanks{Manuscript received: January, 6, 2026; Revised March, 26, 2026; Accepted April, 22, 2026.}
\thanks{This paper was recommended for publication by Editor M. Ani Hsieh upon evaluation of the Associate Editor and Reviewers’ comments. This work was partially supported by the JST's Next-Generation Challenging Research Program JPMJSP2106 and the JAXA, Space Strategy Fund (JPJXSSF24MS09003). Code: https://github.com/stateofyuta-y-cl/NODA.}%
\thanks{$^{1}$Yuta Takahashi is with the Department of Mechanical Engineering, Institute of Science Tokyo, Tokyo, Japan. He is with Satellite Research and Development, Interstellar Technologies Inc., Hokkaido, Japan {\tt\footnotesize stateofyuta@gmail.com}}%
%\thanks{$^{2} $Second Author is with School of Engineering, Automation Department, University of Anywhere, Anyland{\tt\footnotesizel second.author@papercept.net}}%
%\thanks{$^{1}$Ph.D. Candidate, Department of Mechanical Engineering, Institute of Science Tokyo, Ookayama Meguro, Tokyo 152-8552, Japan,}
%\thanks{$^{2}$Researcher, Satellite Research and Development, Interstellar Technologies Inc., 149-7 Hiroo, Hokkaido 089-2113, Japan}
%\thanks{$^{2}$Hayate Tajima is with the Department of Advanced Energy, The University of Tokyo, Chiba, Japan {\tt\footnotesize liverpool.bayern.0821@gmail.com}}
\thanks{$^{2}$Shin-ichiro Sakai is with the Department of Spacecraft Engineering, Japan Aerospace Exploration Agency, Kanagawa, Japan {\tt\footnotesize sakai@isas.jaxa.jp}}%%
\thanks{Digital Object Identifier (DOI): see top of this page.}
%\end{comment}
}
\markboth{IEEE Robotics and Automation Letters. Preprint Version. Accepted April, 2026 (DOI: 10.1109/LRA.2026.3692064)}
{Takahashi \MakeLowercase{\textit{et al.}}: Neural Power-Optimal Magnetorquer Solution for Multi-Agent Formation and Attitude Control} 
\maketitle
\begin{abstract}
This paper presents a learning-based current calculation framework to achieve power-optimal magnetic-field interaction for multi-agent formation and attitude control. In aerospace engineering, electromagnetic coils are referred to as magnetorquer and used as satellite attitude actuators in Earth's orbit and for long-term formation and attitude control. This study derives a unique, continuous, and power-optimal current solution via sequential convex programming and approximates it using a multilayer perceptron model. The effectiveness of our strategy was demonstrated through numerical simulations and experimental trials on the formation and attitude control.% for three electromagnetic coils
\end{abstract}
\begin{IEEEkeywords}
Multi-Robot Systems, Model Learning for Control, Space Robotics and Automation
%Distributed-space system, multi-agent control, nonconvex programming, supervised learning.
\end{IEEEkeywords}
\input{main_text_NODA}
\appendix
\input{appendix_NODA}
%\begin{comment}
%名前と所属は、原稿、付録、補足資料のどこにも含めないでください。
%人々や資金提供機関への謝辞は、原稿が受理された後にのみ記載してください。
\section*{Acknowledgments}%would like to 
\noindent
The authors thank Associate Prof. Yoichi Tomioka at the University of Aizu and Hayate Tajima at The University of Tokyo for technical discussions, Atsuki Ochi at Interstellar Technologies Inc. for assistance with controller gain tuning, and the anonymous reviewers for their valuable comments.
%\end{comment}
%\appendix
%\subsection{Proof of Lemma~\ref{NODA_lemma_reconstruction_phase_vector_fram_X}: Reconstructing Phase $\bm\psi$ from $\mathfrak{X}$}\label{NODA_proof_reconstruction_phase_vector_fram_X}
%\begin{thebibliography}{1}
%\bibliographystyle{IEEEtran}
%\end{thebibliography}
\bibliographystyle{IEEEtran}
\bibliography{references}
\end{document}

%% file: main_text_NODA.tex
\section{Introduction}
\IEEEPARstart{T}{he} magnetic field provides a sophisticated control system used in medicine, biomimetic robotics, and aerospace. Typically, magnetic fields are generated by coils that wind conductive wires around magnetic materials or an air core. The current passing through these conductors generates a magnetic field. Its interaction with the surrounding magnets provides position and orientation information while generating electromagnetic forces and torques. In the medical field, magnetic resonance imaging is a well-known technique that uses these fields to produce non-invasive images of internal structures. In biomedical robotics, these fields can be employed to steer miniature robots inside a human body, enabling targeted drug delivery and minimally invasive procedures \cite{ze2022soft}. The aerospace community originally utilized this actuation in the geomagnetic field as a satellite attitude actuator \cite{sakai2008design}.% and applied satellite state estimation \cite{%nurge2016satellite,shibata2024relative}. 
% through the interaction between onboard magnetic torquers and Earth’s geomagnetic field.
%They have referred to the coils as magnetorquer (MTQ) and recently utilized MTQ in micro-gravity for satellite docking \cite{tajima2023study}, %separation \cite{Inamori_InOrb}, along with navigation %the vibration isolator \cite{Shibata}, debris remover \cite{}, and inter-satellite state estimation \cite{nurge2016satellite}.
\begin{figure}[!t]
\centering
%\begin{minipage}[b]{1\FigWidth}\subfloat[Learning overview of Neural power-Optimal Dipole Allocation (NODA).]{
\includegraphics[width=1\FigWidth]{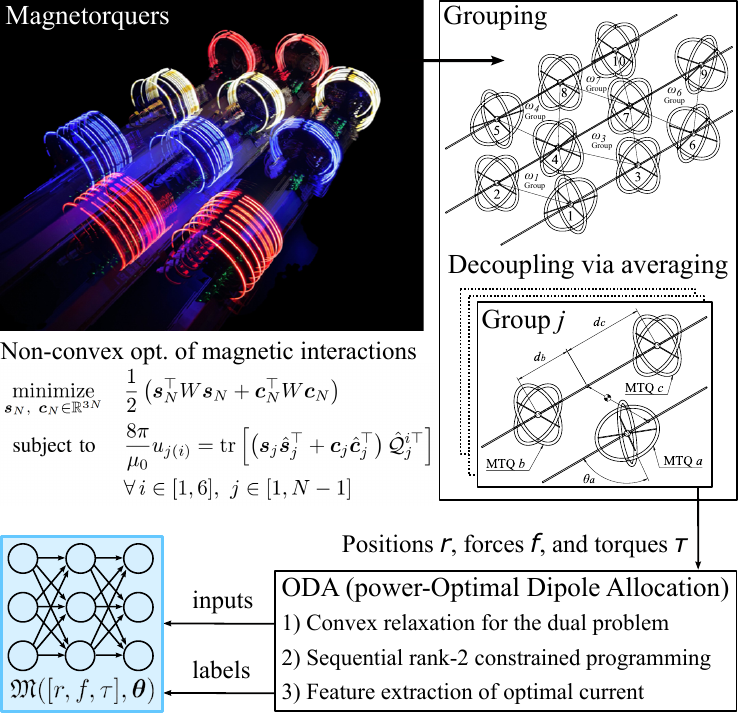}
%}\end{minipage}
%\captionsetup{width=\FigWidth}
\caption{Offline supervised learning of power-optimal dipole allocation model NODA for magnetically actuated multi-agent control.}%
\label{NODA_model_overview}
\end{figure}
%\subsection{\textcolor{black}{Related Work}}

Multi-agent control methodologies using magnetorquers (MTQs) have been developed for the aerospace community. They use magnetic interactions to control the formation and attitudes of satellites. This actuation originally suffered from underactuation for six degrees-of-freedom (DoF) control for $N$ systems \cite{takahashi2022kinematics,takahashi2025noda_mmh}. To deal with this, previous work considered time-integrated current control to extend the controllability for 6-DoF swarm control. Their examples include a time-scheduled switching topology \cite{ramirez2010new} %ramirez2010new
and alternating current (AC) modulation \cite{takahashi2022kinematics}, which controls the amplitude and phase of the AC. The advantage of AC modulation lies in the decoupling of magnetic interactions from different frequencies, including a geomagnetic field that enables decentralized control \cite{takahashi2025noda_mmh}. This actuation provides promising applications for proximity operations \cite{takahashi2026certified} and large distributed space structures \cite{shim2025feasibility,takahashi2025distance}.%They AC method, i.e., they can output arbitrary control force and torque for an arbitrary number of agents. %drives MTQ with the direct current (DC) %\cite{Fabacher,Kwon,Fan,Huang2,Ahsun,Elias,Schweighart,Ramirez-Riberos,Sakaguchi}  %\cite{abbasi2022decentralized,porter2014demonstration,Sakai,Ayyad,Zhang,Youngquist,nurge2016satellite,sunny2019single,Kaneda}.  %The AC method provides higher functionality than DC one: %\cite{Ahsun,Ayyad,Zhang,Kaneda,Sakai}.  %\cite{Schweighart,Fan,Ramirez-Riberos,Ayyad,Huang2} %, the evaluation function $\|\tau\|$ is optimized. %Spacecraft Formation Flying is a concept of controlling large-scale small satellites (0.1-2kg) in close range (0.1-2m) to form large space systems that are not feasible by monolithic space and ground assets (5m-1km). %mounted on small satellites MTQ femto \cite{Hu_Development}. and provides the proof of concepts of formation control by MTQ multiple experiments \cite{alvarez2021multi,hariri2018vision,porter2014demonstration,Sakai_Electromagnetic,sunny2019single,nurge2016satellite}. %位置を制御するformation control用としても使われている．EMFFの位置制御の実験らのリスト．
%\subsection{\textcolor{black}{Contributions and Paper Organization}}
%for three or more satellites, 

Their drawback is that the solution of a computationally intensive optimization problem is required to derive the currents for coordinated magnetic-field interaction, which we refer to as ``dipole allocation.'' This problem contains inherent non-convex constraints \cite{takahashi2022kinematics}, rendering it an NP-hard non-convex quadratically constrained quadratic program (QCQP). As the problem dimensions increase, finding the global optimal solution becomes computationally expensive. Prior studies have explored several computational strategies. General optimization techniques 
%such as the homotopy (continuation) method \cite{morgan2009solving} 
have been applied to derive all optimal solutions \cite{schweighart2006electromagnetic} or track a viable solution using the solution from a previous step \cite{fabacher2017guidance,takahashi2022kinematics}. 
%We avoid using a global solver because it is generally computationally prohibitive for QCQP.
The closed-form Jacobian and Hessian expressions support efficient calculations as specialized methods for this system \cite{abbott2017computing}. For the two-satellite force control case, a unique global minimum power solution was analytically derived in a closed form \cite{abbasi2022decentralized}. However, space-qualified processors offer limited computational capabilities, rendering global solvers impractical. Local solvers provide no performance guarantees, and these locally optimal dipole moment solutions can be discontinuous because the solvers can jump between distinct local minima \cite{takahashi2022kinematics}. This can result in impulsive input and high-frequency disturbances. Moreover, recent applications \cite{takahashi2025distance,shim2025feasibility} assume simultaneous high-precision control of both position and attitude without additional actuators and require repeated computation of the current. These require the optimal dipole solutions to be computed using a significantly reduced computational burden. \textcolor{black}{Moreover, the major challenge arises from the mutual coupling with the other agents, which greatly increases the complexity of the planning and optimization problem.}

To this end, we derive a learning-based approximation model called $\underline{\mathrm{N}}$eural power-$\underline{\mathrm{O}}$ptimal $\underline{\mathrm{D}}$ipole $\underline{\mathrm{A}}$llocation (NODA) for real-time optimal dipole allocation. We also present a systematic multileader-based decentralized control strategy to improve scalability and reduce the sample region. The remainder of this paper is organized as follows: Section~\ref{NODA_Preliminaries} introduces the MTQ system for 6DoF swarm control. Section~\ref{NODA_NODA_subsection} derives a systematic decoupling approach focusing on a small group. In Section~\ref{NODA_Problem_Formulation}, we formulate the optimal dipole allocation problem with a proof of its control-induced disturbance reduction and baseline allocation. Section~\ref{NODA_convex_allocation} presents the main optimization strategy for deriving a continuous, optimal dipole solution that can be approximated by a multilayer perceptron model, as in NODA. Numerical and experimental evaluations validate the effectiveness of the proposed methods and trained model in Section~\ref{NODA_Experimental_Validation}. Section~\ref{NODA_Conclusion} concludes the paper. \textcolor{black}{In this study, the grouping structure is assumed to be fixed and given by previous studies, e.g., centralized optimization \cite{clark2012leader} %distributed strategies based on initial grouping information \cite{schwab2021distributed}
and heuristic approaches \cite{takahashi2025scalable}. The symbols $\otimes$, $\odot$, and $\oslash$ denote the Kronecker product, the Hadamard product, and Hadamard division, respectively. $C^{{B}/A}\in\mathbb{R}^{3\times 3}$ is a coordinate transformation matrix from an arbitrary frame $\mathcal{A}$ to another one $\mathcal{B}$. Arbitrary vector $\bm{x}$ satisfies $\bm{x}=\{\bm b\}^\top x^{b}=\{\bm b\}^\top C^{B/A}x^{a}$ where $\{\bm b\}$ and $x^{b}$ are the basis vectors and the component of $\bm{x}$ in $\mathcal{B}$. All computations use an AMD Ryzen Threadripper 7980X and 64 GB RAM.
}% (Windows 11 Pro, 64-bit).} 
%These centralized controls face scalability and reliability issues due to a high-end communication system, time delays, and computational loads as satellites increase. Distributed strategy generally lacks global superiority and encounters reliability challenges from nearby disturbances. Decentralized Control Strategy with Switching Topology \cite{takahashi2021simultaneous} Hence, control strategies for distributed systems using MTQ should consider continuous trajectory control based on distance constraints and low-current computation, while accounting for magnetic interactions. 
\section{Preliminaries}
\label{NODA_Preliminaries}
%: Orbit and Attitude Control and Electromagnetic System}
%This section summarizes the simultaneous control strategy for the electromagnetic force and torque of $ n$ agents.% and the learning-based functional approximation techniques. 
%\textcolor{black}{This subsection presents the magnetic control model.}
\subsection{Magnetic Interaction Approximation Model}
\label{NODA_EMFF}
We first \textcolor{black}{introduce} the dipole approximation of the magnetic field $\bm{B}_k\left(\bm{\mu}_k, \bm{r}_{j k}\right)$, ``far-field'' model \cite{schweighart2006electromagnetic}. %for large-distance conditions. %, which is accurate if the relative distance exceeds twice the diameter of the coil. 
We denote the $j$th magnetic moment as 
%$${\mu}_{j}(t)=N_t c(t) A {n}\in\mathbb{R}^{3}$$ 
$\bm\mu_{j}=N_t Ac(1+\gamma
%\frac{\left(\mu_r-1\right)}{\left(1-N_d+\mu_r N_d\right)}
)\bm{n}\in\mathbb{R}^3$ 
%Because the ratio of length versus radius is so important it is assumed it can be kept constant in this case. This may not be a valid assumption depending on the difficulty of producing/obtaining the material. It may be necessary to accept a different thickness.
where $N_t$ is the number of coil turns, $c$ is the current
strength, $A$ is the area enclosed by the coil, $\bm{n}$ is the unit vector perpendicular to the coil plane, and $\gamma$ is the iron–core ratio
%$l$ is the length of the coil, $N_d={4(\ln({l}/{r})-1)}/({({l}/{r})^2-4 \ln({l}/{r})})$ 
\cite{%bellini2013magnetic,
takahashi2025noda_mmh}.
%\begin{equation}
%\label{NODA_magnetic_filed}  
%\begin{aligned}
%B_k({\mu}_k, {r}_{j\leftarrow k})=\frac{\mu_0}{4 \pi d_{j\leftarrow k}^3}\left(3M_k\mathsf{e}_r -{\mu}_k\right)
%\end{aligned}
%\end{equation}
This derives force $\bm{f}_{j\leftarrow k}\left(\bm{\mu}_{j},\bm{\mu}_{k}, \bm{r}_{j\leftarrow k}\right)=\nabla(\bm{\mu}_j \cdot \bm{B}_k)$ and torque $\bm{\tau}_{j\leftarrow k}\left(\bm{\mu}_{j},\bm{\mu}_{k}, \bm{r}_{j\leftarrow k}\right)$ exerted on the $j$th agent by the $k$th one:
\begin{equation}
%\label{NODA_basic_formulation}
\begin{aligned}
\bm{f}_{j\leftarrow k}&=\frac{3 \mu_0/4\pi}{ d_{j\leftarrow k}^4}\left[\frac{{\bm{\mu}_k \cdot \bm{\mu}_j}-5 M_k M_j}{ d_{j\leftarrow k} }{\bm{r}_{j\leftarrow k}}+M_k \bm{\mu}_j+M_j\bm{\mu}_k\right]\\
\bm{\tau}_{j\leftarrow k}&=\bm{\mu}_j \times \bm{B}_k=\frac{\mu_0/4 \pi}{d_{j\leftarrow k}^3}\bm{\mu}_j \times \left[\frac{3M_k}{d_{j\leftarrow k}}{\bm{r}_{j\leftarrow k}}-\bm{\mu}_k\right],
%B({\mu}_k, {r}_{j k})=\frac{\mu_0}{4 \pi}\left(\frac{3 {r}_{j k}\left({\mu}_k \cdot {r}_{j k}\right)}{\left\|{r}_{j k}\right\|^5}-\frac{{\mu}_k}{\left\|{r}_{j k}\right\|^3}\right)
\end{aligned}
\end{equation}
where $M_k=\frac{{\mu}_k \cdot {{r}_{j\leftarrow k}}}{d_{j\leftarrow k}}$ and $d_{j\leftarrow k}=\|\bm{r}_{j\leftarrow k}\|$. 
%We can also use exact model \cite{schweighart2006electromagnetic} or its learning-based approximation \cite{takahashi2026certified}.
%Note that this torque model is used often to realize cross-product attitude controller $\tau_{MTQ}=-[B_{e}]_{\times}d_{\mathrm{DC}}$ with geomagnetic field $B_{e}$. 
%where $M_k({\mu}_k,\mathsf{e}_r)={\mu}_k \cdot \mathsf{e}_r$ and $M_j({\mu}_j,\mathsf{e}_r)={\mu}_j \cdot \mathsf{e}_r$.%$d=\left\|{r}_{j k}\right\|$ and $\mathsf{e}_r={{r}_{j k}}/{d}$.
\subsection{Time-Integrated 6-DoF MTQ Control}% for 6-DoF Magnetorquer Control Under Nonholonomy}
\label{NODA_Kinematics_Control}
%\subsection{Alternating Current Magnetorquer Control and ``Kinematics'' for 6-DoF Control \cite{takahashi2022kinematics}}
%\label{NODA_EMFF}
We introduce the AC modulation for 6-DoF control. Let us assume the $j$th agent drives sinusoidal currents as follows:
\begin{equation}
\label{NODA_time-varying-dipole}
%\begin{aligned}
%{\mu}_{j}(t) ={\mu}_{\mathrm{DC}j}+{s}_j \sin (\omega_{j} t)+{c}_j \cos (\omega_{j} t)
\bm{\mu}_{j(t)}
\triangleq\bm{\overline{\mu}}_j \textcolor{black}{\odot}\sin (\omega_{j} t+\bm\psi_j)\triangleq\bm{s}_j \sin (\omega_{j} t)+\bm{c}_j \cos (\omega_{j} t),
%\end{aligned}
\end{equation}
%The total $2\omega_f$ disturbance of electromagnetic force and torque exerted on the $j$th satellite by the entire system is ${f,\tau_{j(2\omega_f)}}=\sum_{k=1}^{n}{f,\tau}_{2\omega_f}(t)$. 
where the amplitudes $\bm{\overline{\mu}}_{j,k}\in\mathbb{R}^3$, angular frequency $\omega_{j}\in\mathbb{R}$, phase $\textcolor{black}{\bm\psi_j}\in\mathbb{R}^3$, and sine and cosine amplitudes $\bm{s}_j,\bm{c}_j\in\mathbb{R}^{3}$. We define the first-order averaged input from the neighbor as $\bm{\overline{u}}_{j}\triangleq\sum_{k\neq j} \bm{\overline{u}}_{j\leftarrow k}$, where $\bm{\overline{u}}_{j\leftarrow k}=[\bm{\overline{f}}_{j\leftarrow k};\bm{\overline{\tau}}_{j\leftarrow k}]$:
%$u_{j\leftarrow k}^{\mathrm{avg}}\triangleq \int_{T}{u}_{j\leftarrow k}({\mu}_{j,k(\tau)}, {r}_{j\rightarrow k})\frac{\mathrm{d}\tau}{T}$
\begin{equation}
\label{NODA_average_far_field_model}
\begin{aligned}
%\bm{\overline{u}}_{j\leftarrow k}&=
&\begin{bmatrix}
\bm{\overline{f}}_{j\leftarrow k}\\
\bm{\overline{\tau}}_{j\leftarrow k}
\end{bmatrix}
%=\begin{bmatrix}f_{j\leftarrow k}^{\mathrm{avg}}\\\tau_{j\leftarrow k}^{\mathrm{avg}}\end{bmatrix}
\triangleq \int_{T}\sum_{k\neq j}\begin{bmatrix}\bm{f}_{j\leftarrow k}\left(\bm{\mu}_{k}(\tau),\bm{\mu}_{j}(\tau), \bm{r}_{j\leftarrow k}\right)\\\bm{\tau}_{j\leftarrow k}\left(\bm{\mu}_{k}(\tau),\bm{\mu}_{j}(\tau), \bm{r}_{j \leftarrow k}\right)\end{bmatrix}\frac{\mathrm{d}\tau}{T}\\
=&\frac{1}{2}
\begin{bmatrix}    
\bm{f}_{j\leftarrow k}(\bm{s_{j}},\bm{s_{k}},\bm{r}_{j\leftarrow k})+\bm{f}_{j\leftarrow k}(\bm{c_{j}},\bm{c_{k}},\bm{r}_{j\leftarrow k})\\
\bm{\tau}_{j\leftarrow k}(\bm{s_{j}},\bm{s_{k}}, \bm{r}_{j\leftarrow k})+\bm{\tau}_{j\leftarrow k}(\bm{c_{j}},\bm{c_{k}},\bm{r}_{j\leftarrow k})
\end{bmatrix}
\end{aligned}
\end{equation}
where this shows that the agent does not interact with other frequency effects in the first-order averaged dynamics \cite{takahashi2022kinematics}, i.e., $\int \bm{u}_{j\leftarrow k}\ \mathrm{d}t\approx 0$ if $\omega_k\neq \omega_j$. This modulation increases the number of variables to realize 6-DoF control for an arbitrary number of agents \cite{takahashi2022kinematics}. We assume all $n$ agents have 3-axis MTQs and $m\in[1,n]$ agents are equipped with 3-axis RWs. Their control holds the angular momentum \cite{takahashi2022kinematics}
\begin{equation}
\label{NODA_angular_momentum_conservation}
%\begin{aligned}
%&\quad\sum_{j=1}^n\left(m_j\left({r}_j-{r}_1\right) \times \frac{\mathrm{d}{r}_j}{\mathrm{d} t}+{I}_j \cdot {\omega}_j\right)+\sum_{j=1}^m {h}_j={L}\\
%\Leftrightarrow &\quad
%&\sum_{j=1}^n(m_j\bm{r}_j\times\bm{\dot{r}}_j+\bm{I}_j \cdot \bm{\omega}_j)+\sum_{j=1}^m \bm{h}_j={L}\ \Leftrightarrow\\
A_{(n,m)}\bm\zeta \triangleq
\begin{bmatrix}
&m_1 [r_{1}^i]_\times, \ldots, m_n [r_{n}^i]_\times,\\
&C^{I/ B_1} J_1, \ldots, C^{I/ B_n} J_n,\\
&C^{I/B_1}, \ldots, C^{I/ B_{m}}
\end{bmatrix}
\begin{bmatrix}
\dot{r}^{a}\\
\omega^{b}\\
\xi^{b}
\end{bmatrix}
=0,
%\end{aligned}    
\end{equation}
where $A_{(n,m)}\in \mathbb{R}^{3 \times(6 n+3m)}$ and the states are $\bm\zeta\in \mathbb{R}^{6 n+3 m}$. A previous study derived the commands $\bm{u}_c\in\mathbb{R}^{6 n+3 m}$ for the submanifold stabilization of magnetically actuated swarms \cite{takahashi2022kinematics}.
\begin{lemma}[Kinematics Control \cite{takahashi2022kinematics}]
\label{NODA_theorem_experimental_controller}
%$S_{(n,m)}\in \mathbb{R}^{(6 n+3 m)\times (6 n+3 m-3)}$
Let $S_{(n,m)}$ be the tangent space of angular momentum conservation $S_{(n,m)}$, e.g.,
\begin{equation}
S_{(n,m)}=
\begin{bmatrix}
E_{(6n+3m-3)}\\
-C^{B_{m}/I}A_{(n,m)[:,1:\mathrm{end-}3]}
\end{bmatrix}%\quad\mathrm{s.t.}\quad A_{(n,m)}S_{(n,m)}=0
\end{equation}
Then, the magnetic interaction can realize $u_c=[f_c^{a};\tau_c^{b};\dot{h}_{c}^{b}]=B_{(n,m)}^{-1}M_{(n,m)} S_{(n,m)} \mathfrak{u}_c$ where the arbitrary auxiliary commands are $\mathfrak{u}_c\in\mathbb{R}^{6 n+3 m-3}$, the inertia matrix is $M_{(n,m)}$, %$=\mathrm{Diag}([M_p,M_\alpha,E_{3m}])$,
and the input matrix is $B_{(n,m)}\in \mathbb{R}^{(6 n+3 m) \times(6 n+3 m)}$ in \cite{takahashi2022kinematics}.
%$$
\end{lemma}
%where $A_{(n,m)}$ is from (\ref{NODA_angular_momentum_conservation}) and $M, B_{(n,m)}$ is from Theorem~\ref{NODA_theorem_experimental_controller}. 
%\begin{proof}See the appendix of \cite{takahashi2022kinematics}.\end{proof} 
\section{Problem Formulation}% of the Power-Optimal Dipole Allocation with Baseline Allocation}% for Approximation}
\label{NODA_Problem_Formulation}
This section formulates the power-optimal dipole allocation problem. We present a multileader-based allocation strategy that decentralizes large-scale agents into smaller groups and derive two allocation formulations for a single group.%: the baseline non-optimal allocation and the optimization-based allocation.
\subsection{Multileader-based Dipole Allocation Strategy}% for Decentralized Control}
\label{NODA_NODA_subsection}
\label{NODA_Guidance_Strategy}
We define $n$ agent nodes as $\mathcal{V}=\{1,\ldots,n\}$ and the edge set as $\mathcal{E} \subseteq \mathcal{V}\times \mathcal{V}$, where edge $(j, k)\in\mathcal{E}$ denotes that the $k$th node \textcolor{black}{obtains} some information from the $j$th node. We define a directed multileader graph $\mathcal{G}^{\mathfrak{ml}}(\mathcal{V}, \mathcal{E}^{\mathfrak{ml}}=\bigcap_j \mathcal{E}_j^{\mathfrak{ml}})$ and assign specific agents to the leader subset $\mathcal{V}_l\textcolor{black}{\subseteq}\mathcal{V}$ with a unique frequency $\omega_{k\in\mathcal{V}_l}$. Its edges $(j, k)\in\mathcal{E}_j^{\mathfrak{ml}}$ describe the hierarchical relationships of $j\in\mathcal{V}_l$ and $k\in\mathcal{V}$, i.e., 
%the $k$th agent following the $j$th one. %We label $j_{\in\mathcal{V}_l}$th leader satellite group as $\mathfrak{g}_{j}$: %$$\mathfrak{g}_{j}\triangleq\{j, k\in \mathcal{V}\  |\ (k, j)\in\mathcal{E}_j^{\mathfrak{ml}} \},\quad\mathcal{E}^{\mathfrak{ml}}\triangleq\mathcal{E}_1^{\mathfrak{ml}} \cap\ldots \cap \mathcal{E}_{n}^{\mathfrak{ml}}$$ %$\mathfrak{g}_{j}=j$ and 
%Note that $\mathcal{E}_{j}^{\mathfrak{ml}}=\{\}$ indicates $j$th leader agent does not have followers. of the $j$th leader 
the $k$th agent drives the $\omega_{j\in\mathcal{V}_l}$ current. This extends the dipole moment in (\ref{NODA_time-varying-dipole}) into the multi-frequency one:
%\begin{equation}
%\label{NODA_time-varying-dipole}
%\begin{aligned}
%{\mu}_{j}(t) ={\mu}_{\mathrm{DC}j}+{s}_j \sin (\omega_{j} t)+{c}_j \cos (\omega_{j} t)
%\end{aligned}
%\end{equation}
\begin{equation}
\label{NODA_multi_frequency_dipole_moment}
%\begin{aligned}
\bm{\mu}_{k}(t) =%\boldsymbol{\mu}_{\mathrm{DC}j}+
\sum_{j\in\mathcal{V}_l}%\sum_{(j,k)\in\mathcal{E}_j}
(\bm{s_{k[j]}} \sin (\omega_{j} t)+\bm{c_{k[j]}} \cos (\omega_{j} t)),\\
%&\vdots\\
%+{s_j} \sin \left(\omega_{fk} t\right)&+{c_j} \cos \left(\omega_{fk} t\right).
%\end{aligned}
\end{equation}
where $\bm{s}_{k[j]},\bm{c}_{k[j]}\in\mathbb{R}^{3}$ are the $k$th sine and cosine amplitude in the $j_{\in\mathcal{V}_l}$th leader group. One cycle $T$ and each $\omega_{j}$  should be derived to cancel out the coupling between different frequencies and hold the first-order approximation in (\ref{NODA_average_far_field_model}) as follows:%\cite{takahashi2022kinematics}
\begin{equation}
\label{NODA_AC_cycle}
\small{\forall \ 1 \leq i<j \leq n},\quad
T\triangleq{\mathrm{lcm}\left(\tfrac{2\pi k_{\mathrm{int}}}{2\omega_{i}}, \tfrac{2 \pi k_{\mathrm{int}}}{\left|\omega_{i}-\omega_{j}\right|}, \tfrac{2 \pi k_{\mathrm{int}}}{\omega_{i}+\omega_{j}}\right)}/{k_{\mathrm{int}}}
\end{equation}
where $\mathrm{lcm}(\cdot)$ is the least common multiple and the coefficient $k_{\mathrm{int}}$ ensures the elements of $\mathrm{lcm}(\cdot)$ are integers. For a user-defined $T_{\mathrm{max}}$, the necessary conditions for $T\leq T_{\mathrm{max}}$ are
\begin{equation}
\omega_{i}\geq \frac{\pi}{T_{\mathrm{max}}},\quad |\omega_{i}-\omega_{j}|\geq \frac{2\pi}{T_{\mathrm{max}}},\quad \omega_{i}+\omega_{j}\geq \frac{2\pi}{T_{\mathrm{max}}},
\end{equation}
such that the elements in (\ref{NODA_AC_cycle}) satisfy $T\leq T_{\mathrm{max}}$. %, and $i,j$ is arbitrary index that satisfies $\ 1 \leq i<j \leq n$.
\begin{remark}
\textcolor{black}{Note that the magnetic interaction is naturally decentralized among groups that are sufficiently far apart.
%, since the induced force and torque decay proportionally to the inverse fourth and third powers of distance, respectively. 
Therefore, a minimum number of distinct AC frequencies is equal to the maximum number of groups that can coexist within the effective interaction range of any given satellite.}
One easy way to select $\omega_{i}$ for a given $T_{\mathrm{max}}$ is
\begin{equation}
\omega_{i}=\frac{2\pi i}{T_{\mathrm{max}}}\ \mathrm{s.t.}\ T=\frac{ \mathrm{lcm}\left(\frac{ k_{\mathrm{int}}}{2i},\ \frac{k_{\mathrm{int}}}{|i-j|},\ \frac{ k_{\mathrm{int}}}{i+j}\right)} {k_{\mathrm{int}}/T_{\mathrm{max}}}=T_{\mathrm{max}}.
\end{equation}
%In this case, the current should be updated based on the maximum frequency $(2n+1)\times ({2\pi}/{T_{\mathrm{max}}})$ for an averaging. 
\end{remark}
%Strict Global Power-Optimal Dipole Allocation by Convex-Optimization for 6-DoF Magnetorquer Control
%Using the multileader-based decentralization introduced in the previous section, t
\subsection{Baseline Dipole Allocation for Decentralized Control}
%Motivated by prior work \cite{abbasi2022decentralized}, 
We derive the baseline decentralized dipole allocation using an inverse matrix, motivated by prior work \cite{abbasi2022decentralized}.
First, we simplify the time-averaged far-field magnetic interaction model as a bilinear polynomial formulation. For given arbitrary vectors $v{^a},w{^a}\in\mathbb{R}^3$, we newly define the coordinates such that its $x$-axis aligns with $v{^a}$, i.e., $\mathsf{e}_x=\mathrm{nor}({v{^a}})$, its $y$-axis is orthogonal to $v{^a}$, and $w{^a}$ satisfies $\mathsf{e}_y=\mathrm{nor}(S({v{^a}})w{^a})$. This yields the following coordinate transformation matrix:% $\mathcal{C}_{({^a}v,{^a}w)}\in\mathbb{R}^{3\times 3}$ as
\begin{equation}
\label{NODA_new_frame}
%\begin{aligned}
\mathcal{C}{(v{^a},w{^a})}=
      \begin{bmatrix}
\mathsf{e}_x=\frac{v{^a}}{\|v{^a}\|},\mathsf{e}_y=
\frac{[{v{^a}}]_\times w{^a}}{\|[{v{^a}}]_\times w{^a}\|},\mathsf{e}_x\times \mathsf{e}_y
     \end{bmatrix}.%,\quad \mathsf{e}_y=\mathrm{nor}(S({v{^a}})w{^a})
%\end{aligned}
\end{equation}
%where $\mathsf{e}_y=\mathrm{nor}(S({v{^a}})w{^a})$. 
\begin{definition}
\label{NODA_C_LOS_new}
The line-of-sight frame $\{\mathcal{LOS}_{j\leftarrow k}\}$, fixed for the $k$th agent so that it can see the $j$th one, is defined by (\ref{NODA_new_frame}) such that its coordinate transformation matrix $C^{O/L_{j\leftarrow k}}\in\mathbb{R}^{3\times 3}$ is
\begin{equation}
%\begin{aligned}
%&C^{O/LOS}_{jk}=\begin{bmatrix}\mathsf{e}_x\triangleq\mathrm{nor}({{^o}r_{jk}}),\mathsf{e}_y,\mathsf{e}_x\times \mathsf{e}_y\end{bmatrix}\in\mathbb{R}^3\\
C^{A/L_{j\leftarrow k}}=\mathcal{C}(r_{j\leftarrow k}^a,\ S(f^a_{j})r^a_{j\leftarrow k}),
%\end{aligned}
\end{equation}
where $x$-axis in $\mathcal{LOS}$ is aligned with $r_{jk}$ and $f_{j}$ remains in the $x$-$y$ plane in $\mathcal{LOS}$, i.e., $r^{l_{j\leftarrow k}}_{j\leftarrow k}=[\|r^{a}_{j\leftarrow k}\|;0;0]$ and $f^{l_{j\leftarrow k}}_{j(3)}=0$. 
\end{definition}
%\begin{definition}
%\label{NODA_los_interaction_definition}
\noindent
The frame in Definition~\ref{NODA_C_LOS_new} yields the bilinear polynomial formulation in (\ref{NODA_average_far_field_model}) with the constant matrix 
\begin{equation}
\label{NODA_time_averaged_LOS}
%\begin{aligned}
u_{j\leftarrow k}^{a}\approx \overline{u}_{j\leftarrow k}^{a}=
\frac{1}{2}\frac{\mu_0}{4\pi}Q_{{j\leftarrow k}}
\left (
{s_{k}^{a}}
\otimes
{s_{j}^{a}}
+
{c_{k}^{a}}
\otimes
{c_{j}^{a}}
\right ),
%&=\frac{\mu_0}{8\pi}\mathrm{tr}[s_{j}^{a}({s_{k}^{a\top}}\mathcal{Q}_{j\leftarrow k[i]}^\top) +c_{j}^{a}({c_{k}^{a\top}} \mathcal{Q}_{j\leftarrow k[i]}^\top)]\\
%\end{aligned}
\end{equation}
%このaとbをlとfに変更する。
where %$u^{los}_{L\leftarrow F}=[{f^{los}_{L\leftarrow F}};{\tau^{los}_{L\leftarrow F}}]\in \mathbb{R}^{6}$, 
we define $d_{j\leftarrow k}\triangleq\|\bm{r}_{j\leftarrow k}\|$ and $Q_{{j\leftarrow k}}\in \mathbb{R}^{6\times 9}$ is
\begin{equation}
%\label{NODA_sec2::eq18}
%\left\{
\begin{aligned}
&Q_{{j\leftarrow k}}= 
\begin{bmatrix}
C^{A/L_{j\leftarrow k}}\Psi_{f({d_{j\leftarrow k}})}\\
C^{A/L_{j\leftarrow k}}\Psi_{\tau({d_{j\leftarrow k}})}
\end{bmatrix}
(C^{L_{j\leftarrow k}/A}\otimes C^{L_{j\leftarrow k}/A})\\
&\left\{
\begin{aligned}
&\Psi_f
%({d_{j\leftarrow k}})
=
%r_{\mathrm{sgn}}
\frac{1}{{d_{j\leftarrow k}^4}}
{\small\begin{bmatrix}
%-6,0_3,3,0_3,3\\0,3,0,3,0_5\\0_2,3,0_3,3,0_2\\
-6&0&0&0&3&0&0&0&3\\
0&3&0&3&0&0&0&0&0\\
0&0&3&0&0&0&3&0&0
%\end{bmatrix}&\begin{bmatrix}
%0&3&0\\3&0&0\\0&0&0
%\end{bmatrix}&\begin{bmatrix}
%0&0&3\\0&0&0\\3&0&0
\end{bmatrix}}
%d_{j\leftarrow k}\begin{bmatrix}
%0_5,1,0,-1,0\\
%0_2,2,0_3,1,0_2\\
%0,-2,0,-1,0_5\\
%0&0&0&0&0&1&0&-1&0\\
%0&0&2&0&0&0&1&0&0\\
%0&-2&0&-1&0&0&0&0&0
%\end{bmatrix}
%&\begin{bmatrix}
%\\0&0&0\\-d^2&0&0
%\end{bmatrix}&\begin{bmatrix}
%0&-d^2&0\\d^2&0&0\\0&0&0
%\end{bmatrix}
%\end{bmatrix}
\\
&%\frac{\Psi({d_{j\leftarrow k}})}{{1}/{d_{j\leftarrow k}^4}}
\Psi_\tau
%({d_{j\leftarrow k}})
=
%\begin{bmatrix}
%r_{\mathrm{sgn}}
%\begin{bmatrix}
%-6,0_3,3,0_3,3\\0,3,0,3,0_5\\0_2,3,0_3,3,0_2\\
%-6&0&0&0&3&0&0&0&3\\
%0&3&0&3&0&0&0&0&0\\
%0&0&3&0&0&0&3&0&0\\
%\end{bmatrix}&\begin{bmatrix}
%0&3&0\\3&0&0\\0&0&0
%\end{bmatrix}&\begin{bmatrix}
%0&0&3\\0&0&0\\3&0&0
%\end{bmatrix}\\
\frac{1}{d_{j\leftarrow k}^3}
{\small\begin{bmatrix}
%0_5,1,0,-1,0\\
%0_2,2,0_3,1,0_2\\
%0,-2,0,-1,0_5\\
0&0&0&0&0&1&0&-1&0\\
0&0&2&0&0&0&1&0&0\\
0&-2&0&-1&0&0&0&0&0
\end{bmatrix}}
\end{aligned}
\right.
%&\begin{bmatrix}
%\\0&0&0\\-d^2&0&0
%\end{bmatrix}&\begin{bmatrix}
%0&-d^2&0\\d^2&0&0\\0&0&0
%\end{bmatrix}
%\end{bmatrix}
\end{aligned}.
%\right.
\end{equation}
%where $\sigma_{\max}(\Psi_f)=\sqrt{54}/d_{j\leftarrow k}^4$ and $\sigma_{\max}(\Psi_\tau)=\sqrt{5}/d_{j\leftarrow k}^3$. 
%\end{definition}%$r_{\mathrm{sgn}}=\mathrm{sgn}(r_{j\leftarrow k}(1))=\frac{^ir_{j\leftarrow k}(1)}{d}$.
%\subsection{Baseline: Inverse-based Decentralized Dipole Allocation}
%\label{NODA_Inverse}
\textcolor{black}{Note that the rigorous formulation of the magnetic interaction includes sinusoidal disturbances $\bm{\widetilde{d}}_{j\leftarrow k(t)}\in\mathbb{R}^{6}$ as $\bm{u}_{j\leftarrow k}(t)=\bm{\overline{u}}_{j\leftarrow k}+\bm{\widetilde{d}}_{j\leftarrow k}(t)$ as shown in (\ref{NODA_average_far_field_model}) and we prove in Appendix \ref{NODA_control_induced_disturbance_reduction} that this power-optimal dipole allocation also leads to a reduction in time-integrated control-induced disturbances.} We allocate a dipole for two agents in a decentralized manner. In (\ref{NODA_time_averaged_LOS}), we randomly predetermine one dipole as a constant indexed by 0, e.g., $\{\bm{s}_{F0},\bm{c}_{F0}\in\mathbb{R}^{3}\mid\|\bm{s}_{F0}\|=\|\bm{c}_{F0}\|=1\}$. This derives another one using the inverse matrix calculation
%the nature of the bilinear polynomial:
\begin{equation}
\label{NODA_decentralized_allocation}
\left\{
\begin{aligned}
\begin{bmatrix}
{s_{L}^{los}}\\
{c_{L}^{los}}
\end{bmatrix}&=
\left\{
\frac{\mu_0}{8\pi}
Q_{L\leftarrow F}
\left(\begin{bmatrix}
s_{F0}^{los},c_{F0}^{los}\end{bmatrix} \otimes
   E_3\right)
\right\}^{-1}u_{L\leftarrow F}^{los}\\
\begin{bmatrix}
{s_{F}^{los}}\\
{c_{F}^{los}}
\end{bmatrix}&=
\left\{
\frac{\mu_0}{8\pi}
Q_{L\leftarrow F}
\left(E_3 \otimes
   \begin{bmatrix}
s_{L0}^{los},c_{L0}^{los}\end{bmatrix}\right)
\right\}^{-1}u_{L\leftarrow F}^{los}
\end{aligned}
\right.
\end{equation}
where the necessary conditions for this inverse are noted in Appendix~\ref{Necessary_Conditions_of_Baseline}. We can minimize $\|\bm{\overline{\mu}}_{N}\|^2$ by a normalization:
\begin{equation}
\label{norm_normalization}
\{s_L,c_{L}\}=\frac{\{s_{L0},c_{L0}\}}{\sqrt{\|\{s_{L0},c_{L0}\}\|}},\ \{s_F,c_F\}=\frac{\{s_{F0},c_{F0}\}}{1/{\sqrt{\|\{s_{L0},c_{L0}\}\|}}}.
\end{equation}
We summarize this randomized allocation in Algorithm~\ref{NODA_alg:Inverse_based_dipole_allocation}. The baseline allocation in (\ref{NODA_time_averaged_LOS}) can yield a decentralized strategy that does not require intersatellite communication.
\begin{algorithm}[tb!]
\begin{algorithmic}[1] 
\STATE  \textbf{Inputs: }trial number $n_{\mathrm{rand}}$, ${r}_{L\leftarrow F}\in\mathbb{R}^3,u_{L\leftarrow F}^{los}\in\mathbb{R}^{6}$
\STATE  \textbf{Outputs: }$[s_L;s_F;c_L;c_F]^{(i^*)}\in\mathbb{R}^{12}$%, $W_{\mathrm{power}}^{(i^*)}\in\mathbb{R}$
%\STATE Calculate $u_{L\leftarrow F}^{los}$ and $Q_{j\leftarrow k}$ in Definition~\ref{NODA_los_interaction_definition}
\FOR{$i\in[1,n_{\mathrm{rand}}]$}
%\STATE Generate randomly $\{s_F,c_F\in\mathbb{R}^{3}\mid\|s_F\|=\|c_F\|=1\}$% based on Remark2 \ref{NODA_}
%\STATE Calculate $s_L,c_L$ in (\ref{NODA_decentralized_allocation}) and 
\STATE Derive $[s_L;s_F;c_L;c_F]^{(i)}$ by (\ref{NODA_decentralized_allocation}) and (\ref{norm_normalization})% and $W_{\mathrm{power}}^{(i)}$% by Remark~\ref{NODA_Power_Normalization_Decentralized_Dipole_Allocation}
\ENDFOR
\STATE Choose %$i^*$ %$[s_L;s_F;c_L;c_F]^{(i^*)}$ 
%where 
$i^*=\argmin\ _i \|[s_L;s_F;c_L;c_F]^{(i)}\|^2$
%\STATE Derive $\mu_j^a(t)=\bm{\overline{\mu}}_j^*\sin(\omega_{f}t+\theta_j^*+\theta_0)$
\end{algorithmic}
\caption{Inverse-based dipole allocation based on (\ref{NODA_decentralized_allocation}).}
\label{NODA_alg:Inverse_based_dipole_allocation}
\end{algorithm}
%\begin{remark}[Power Normalization for Baseline Allocation]
%\label{NODA_Power_Normalization_Decentralized_Dipole_Allocation}
%\end{remark}
%We show that the lack of an arbitrary component in LOS falls into the singularity of the inverse matrix.
%$$
%\mu_{j\bot}/ \! /({c}_j\times {s}_j)=\begin{bmatrix}
%\bm{\overline{\mu}}_{j(2)}\mu^{\mathrm{amp}}_{j(3)}\sin(\theta_{j(2)} - \theta_{j(3)})\\
%\mu^{\mathrm{amp}}_{j(3)}\mu^{\mathrm{amp}}_{j(1)}\sin(\theta_{j(3)} - \theta_{j(1)})\\
%\mu^{\mathrm{amp}}_{j(1)}\mu^{\mathrm{amp}}_{j(2)}\sin(\theta_{j(1)} - \theta_{j(2)})
%\end{bmatrix}.
%$$
%$$
%\mu_{F\bot}/ \! /
%\begin{bmatrix}
%\mu^{\mathrm{amp}}_{F(2)}\mu^{\mathrm{amp}}_{F(3)}\sin(\theta_{F(2)} - \theta_{F(3)})\\
%\mu^{\mathrm{amp}}_{F(3)}\mu^{\mathrm{amp}}_{F(1)}\sin\theta_{F(3)}\\
%-\mu^{\mathrm{amp}}_{F(1)}\mu^{\mathrm{amp}}_{F(2)}\sin\theta_{F(2)}
%\end{bmatrix}.
%$$
\subsection{Power-Optimal Dipole Allocation and Lower Power Bound}
%First, %we stack the vectors of the time-varying dipole moments in (\ref{NODA_time-varying-dipole}) as $\mu^{\mathrm{amp}}_N,\bm\psi_N,s_N,c_N\in\mathbb{R}^{3n}$.
\begin{comment}
\begin{equation}
\label{NODA_dipole_vectors}
\mu^{\mathrm{amp}}_N \triangleq 
    \begin{bmatrix}
    \mu^{\mathrm{amp}}_1\\
    \vdots\\
    \mu^{\mathrm{amp}}_n
    \end{bmatrix},\ 
\bm\psi_N \triangleq 
    \begin{bmatrix}
    \bm\psi_1\\
    \vdots\\
    \bm\psi_n
    \end{bmatrix},\     
s_N \triangleq
    \begin{bmatrix}
    s_1\\
    \vdots\\
    s_n
    \end{bmatrix},\ 
c_N \triangleq
    \begin{bmatrix}
    c_1\\
    \vdots\\
    c_n
    \end{bmatrix}
\end{equation}
\end{comment}
%where $\mu^{\mathrm{amp}}_N,\bm\psi_N\in\mathbb{R}^{3n}$.
The baseline allocation is not power-optimal, which motivated us to derive optimal power-consumption solutions that are crucial for limited resources. We define the overall dipole vector ${m}(t)\in\mathbb{R}^{6n}$ for the $j_{\in\mathcal{V}_l}$th leader group as 
\begin{equation}
\label{dipole_moment_vectors_m_tau}
\bm{m}(t)=
\begin{bmatrix}
    \bm{\overline{\mu}}_{N}\odot\sin(\omega_jt+\bm\psi_N)\\
    \bm{\overline{\mu}}_{N}\odot\cos(\omega_jt+\bm\psi_N)
\end{bmatrix}
%{m}(\mu^{\mathrm{amp}}_{N},\bm\psi_N,t)=
=\begin{bmatrix}
    \sin(\omega_jt)\bm{s}_N\\
    \cos(\omega_jt)\bm{c}_N
\end{bmatrix}
\end{equation}
where $\bm{\overline{\mu}}_N,\bm\psi_N,\bm{s}_N,\bm{c}_N\in\mathbb{R}^{3n}$ are stacked vectors of (\ref{NODA_time-varying-dipole}). 
\begin{comment}
For a given command 6-DoF control $\bm{f}_{cj},\bm{\tau}_{cj}$ by an arbitrary controller, the power-optimal dipole solutions $m^{*}$ satisfy the QCQP problem:
\begin{equation}
\label{NODA_basic_formulation}
%\begin{aligned}
\min_{\bm{s}_N,\bm{c}_N\in\mathbb{R}^{3n}}%J_p(m)=
\ \int
%_{t\in[0,T]}
\frac{\|\bm{m}(\tau)\|^2}{2}\frac{\mathrm{d}\tau}{T}\quad\mathrm{s.t.}\quad \bm{u}_{cj}=\bm{\overline{u}}_{j},\quad\forall j_{\in[1,n]}
%\left\{\begin{aligned}{f}_{cj}&={f}_{j(\mathrm{avg})},\\{\tau}_{cj}&={\tau}_{j(\mathrm{avg})},\end{aligned}\right.
%\end{aligned}
\end{equation}
%for $j\in[1,n]$ 
\end{comment}
For a given force and torque command $u_c^{a}=[u_{c1}^a;\ldots;u_{cn}^a]\in\mathbb{R}^{6n}$ with $[u_{cj}^a=\bm{f}_{cj};\bm{\tau}_{cj}]$ by an arbitrary controller, we define the power-optimal dipole allocation problem with an infinite number of solutions \cite{takahashi2022kinematics} and $6n$ equality constraints:%\begin{equation}u^{a}=[u_{1}^a;\ldots;u_{n}^a]\in\mathbb{R}^{6n},\quad u_{j}^{a}=%\frac{\mu_0}{8\pi}\sum_{k\neq j}u_{j\leftarrow k}^{a},\end{equation}
\begin{equation}
\label{NODA_basic_formulation}
%\left\{
\begin{aligned}
&\min_{\bm{s}_N,\bm{c}_N\in\mathbb{R}^{3n}}\ %J(m)=\frac{1}{2}m^\top Wm=
%_{t\in[0,T]}
\int_T%\frac{
\|W^{\tfrac{1}{2}}\bm{m}{(\tau)}\|^2
%}{2}
\frac{\mathrm{d}\tau}{T}%\right)=
=\frac{\bm{s}_N^\top W \bm{s}_N+\bm{c}_N^\top W \bm{c}_N}{2}%\left(
\\
&\mathrm{s.t.\ }
%&\sum_{k\neq j}Q_{d_{jk}}^a ({s_{k}^a}\otimes {s_{j}^a}+{c_{k}^a}\otimes{c_{j}^a})=\frac{8\pi}{\mu_0}u_{j}^{a}\\
%\left\{
%\begin{aligned}
%&\ \ \\
u_{cj(i)}^{a}=\frac{\mu_0}{8\pi}\sum_{k\neq j}Q_{{j\leftarrow  k}(i,:)}^a({s_{k}^a}\otimes{s_{j}^a}+{c_{k}^a}\otimes{c_{j}^a})
%\mathrm{tr}\left[\left(s_{j}^{a}\hat{s}_j^{a\top} +c_{j}^{a}\hat{c}_j^{a\top} \right) {\mathcal{Q}}_{j[i]}^{\top}\right]
%\\
%\end{aligned}
%\right.
\end{aligned}
%\right.
\end{equation}
for $i\in[1,6]$, $j\in[1,n-1]$, and a constant weight matrix $W$. Its associated Lagrangian $L(\bm{m},\bm{\lambda})$ of (\ref{NODA_basic_formulation}) is
\begin{equation}
L\triangleq \tfrac{1}{2}\bm{m}^\top (I_2\otimes P_{\bm{\lambda}}) \bm{m}-
\tfrac{8\pi}{\mu_0}\bm{\lambda}^\top
\begin{bmatrix}
u_{c1}^{a}; \ldots; 
%\vdots\\
u_{c(n-1)}^{a}
\end{bmatrix},
\end{equation}
where the Lagrange multiplier vector $\bm{\lambda}\in\mathbb{R}^{6(n-1)}$ and $P_{\bm{\lambda}}\triangleq I_{3n}+R_{\bm{\lambda}}+R_{\bm{\lambda}}^\top$ and $R_{\bm{\lambda}}\in\mathbb{R}^{3n\times 3n}$ is the following non-symmetric matrix with $\mathrm{vec}({R}_{j\leftarrow k})\triangleq Q^{\top}_{j\leftarrow k}\bm\lambda_j$:
\begin{equation}
%\label{NODA_P_n}
%\begin{aligned}
\small{R_{\bm{\lambda}}\triangleq
\begin{bmatrix}
O_3&{R}_{1\leftarrow 2}&\cdots&{R}_{1\leftarrow (n-1)}&{R}_{1\leftarrow n}\\
{R}_{2\leftarrow 1}&O_3&\cdots
&{R}_{2\leftarrow (n-1)}&{R}_{2\leftarrow n}\\
\vdots&\vdots&\ddots
&\vdots&\vdots
\\
{R}_{(n-1)\leftarrow 1}&{R}_{(n-1)\leftarrow 2}&\cdots
&O_3&{R}_{{(n-1)}\leftarrow n}\\
O_3&O_3&\cdots&O_3&O_3
\end{bmatrix}}
%\end{aligned}
\end{equation}
Note that $P_{{\lambda}}$ is a positive definite matrix; otherwise, there exists an $m$ such that $L(\bm{m},\bm{\lambda})\rightarrow-\infty$. Then, we obtain the lower bound as $J_{\mathrm{dual}}\leq \frac{1}{2}{\|\bm{m}\|^2}$ using 
%Let ${\lambda}^*$ be defined as the optimum solution of 
the following Lagrange dual problem \cite{boyd2004convex}, which is convex and
efficiently solvable:
\begin{comment}
\begin{equation}
\label{NODA_opt2_N}
%\mathcal{OPT}_{\mathrm{DUAL}}\ 
\max_{\lambda \in \mathbb{R}^{6(n-1)}}
%_{{\lambda}\in\mathbb{R}^{6(N-1)}}
\ J_{\mathrm{dual}}=
%\frac{-{\lambda}^\top \hat{u}^{a}}{\mu_0/(8\pi)}
-\frac{8\pi}{\mu_0}{\lambda}^\top
\begin{bmatrix}
u_{c1}^{a};\ldots;
%\vdots\\
u_{c(n-1)}^{a}
\end{bmatrix}\ \mathrm{s.t.}\  P_{{\lambda}}\succeq 0.
\end{equation}
\end{comment}
\begin{equation}
\label{NODA_opt2_N}
J_{\mathrm{dual}}
=\max_{\{\bm{\lambda}_j\in\mathbb{R}^6\}_{j=1}^{n-1}}\ 
-\frac{8\pi}{\mu_0}
\sum_{j=1}^{n-1} \lambda_j^\top u_{cj}^a
\quad \mathrm{s.t.}\quad P_{\bm{\lambda}} \succeq 0 .
\end{equation}
\section{Neural Power-Optimal Dipole Allocation}% for Approximation}
\label{NODA_convex_allocation}
%Strict Global Power-Optimal Dipole Allocation by Convex-Optimization for 6-DoF Magnetorquer Control
This subsection presents the learning-based power-optimal dipole calculation approximation framework for the control of $n$ agents. %\subsection{Multilayer Perceptron Model Approximation}
A multilayer perceptron (MLP) model represents the functional mapping from inputs $\bm{x}$ into outputs $\bm{y}$ such as an ($L$+1)-layer neural network $\bm{y}\approx\bm{\mathcal{F}}(\bm{x}, {\bm\theta}) = 
W^{(L+1)}\phi(\cdots\phi(W^{(1)} \bm{x}+\bm{b}^{(1)})\cdots)+\bm{b}^{(L+1)}$ where the activation function $\phi(\cdot)$ and $\bm\theta$ include weights $\bm\theta_w$ 
%. = W^{(1)},\ldots, W^{(L+1)}$ 
and bias $\bm\theta_b$.% = \bm{b}^{(1)},\ldots, \bm{b}^{(L+1)}$.%, and are trained to minimize a loss function.
\subsection{Minimal Representation of the Dipole Solution}
\label{NODA_minimal_representation_dipole_solution}
%これは合成させると振幅と位相差を変数としていることになる。この位相差以外の任意の時刻からのスタートが解の無限化につながっている。それを無くせばただのnonconvex問題。そこにKTT条件。さらに，冗長な定式なので一つの評価関数に対して必ず無限個の解が存在する．それを示す．
The result by $\bm{m}(t)$ in (\ref{NODA_basic_formulation}) is achieved by $\bm{m}(t+\psi_0)$ with arbitrary $\psi_0$ 
%We rewrite  ${s}_{j}\in\mathbb{R}^3$ and ${c}_{j}\in\mathbb{R}^3$ to express this redundant of the time-varying dipole $\mu_j(t)$: 
\begin{comment}
\begin{equation}
\label{NODA_new_definition_infinite}
%\begin{aligned}
{s}_{j(l)}%(%\mu^{\mathrm{amp}}_N,\bm\psi_N,\psi_0)
=\mu^{\mathrm{amp}}_{j(l)}\cos(%N_c\pi+
\psi_0+\bm\psi_{j(l)}),\quad{c}_{j(l)}%(%\mu^{\mathrm{amp}}_N,\bm\psi_N,\psi_0)
=\mu^{\mathrm{amp}}_{j(l)}\sin(%N_s\pi+
\psi_0+\bm\psi_{j(l)})
%\end{aligned}
\end{equation}
\end{comment}
%for all $j\in[1,n]$ and $l\in\{1,2,3\}$ 
since we rely on the first-order averaged model. Then, we introduce a matrix variable $\mathfrak{X} \in\mathbb{\mathbb{R}}^{3n \times 3n}$
%as follows:(\ref{dipole_moment_vectors_m_tau})
\begin{equation}
\label{definition_mathfrakX}
%\begin{aligned}
\mathfrak{X}\triangleq2\int_T  \bm{m}{(\tau)}\bm{m}{(\tau)}^\top\frac{\mathrm{d}\tau}{T}%=s_N s_N^\top +c_N c_N^\top
=\begin{bmatrix}\bm{s}_N&\bm{c}_N\end{bmatrix}\begin{bmatrix}\bm{s}_N&\bm{c}_N\end{bmatrix}^\top
%\end{aligned}
\end{equation}
where the final equality denotes $\mathrm{rank}(\mathfrak{X})\leq 2$. 
%Although this extends $\bm{s}_N,\bm{c}_N\in\mathbb{R}^{3n}$ to $\mathfrak{X} \in\mathbb{\mathbb{P}}^{3n \times 3n}$, a \textcolor{black}{vector} $\mathtt{x}(\mathfrak{X})\in\mathbb{R}^{9n}$% for a given $\mathfrak{X}$
Although this lifts the number of variables from $3n$ to $9n^2$, it is sufficient, for reconstruction of the original data, to retain a \textcolor{black}{vector} $\bm{\mathtt{x}}(\mathfrak{X})$:
\begin{equation}
\label{NODA_mathfrak_y_reconstructing}
\bm{\mathtt{x}}(\mathfrak{X})=
\begin{bmatrix}
\mathrm{Diag}(\mathfrak{X})^\top & \mathfrak{X}{(:,1)}^\top & \mathfrak{X}{(:,2)}^\top
\end{bmatrix}^\top \in\mathbb{R}^{9n}
\end{equation}
by reconstructing the trigonometric terms of each phase. 
%From this vector, the trigonometric terms of each phase can be reconstructed, thereby recovering the original $3n$-dimensional information. Although this extends the number of variables from $3n$ to $9n^2$, only a \textcolor{black}{vector} $\mathtt{x}(\mathfrak{X})\in\mathbb{R}^{9n}$
%Reconstructing Phase $\bm\psi$ from $\mathfrak{X}$
\begin{lemma}
%[Reconstructing Phase $\bm\psi$ from $\mathfrak{X}$]
\label{NODA_lemma_reconstruction_phase_vector_fram_X}
A given matrix $\mathfrak{X}$ yields $m(t)=[s_N;c_{N}]$:% as follows:
\begin{equation}
\label{NODA_reconstruction_c_s_N}
\left\{
\begin{aligned}
\bm{c}_{N}&\triangleq \bm{\overline{\mu}}_{N}\odot\mathrm{COS}_{\mathfrak{X}(:,1)},\quad\bm{\overline{\mu}}_{N}=\sqrt{\mathrm{Diag}(\mathfrak{X})},\\
\bm\psi_{1(1)}&\triangleq 0,\quad \mathrm{COS}_{\mathfrak{X}}\triangleq\mathfrak{X} \oslash (\bm{\overline{\mu}}_{N}\bm{\overline{\mu}}^{\top}_{N}),\\
\bm{s}_{N}&\triangleq\bm{\overline{\mu}}_{N}\odot\left(\mathrm{sign}(\sin\bm\psi_{2})\sin\bm\psi_N\right),\\
\sin\bm\psi_{x}&\triangleq\tfrac{\mathrm{COS}_{\mathfrak{X}(x,2)}-\mathrm{COS}_{\mathfrak{X}(x,1)}\mathrm{COS}_{\mathfrak{X}(2,1)}}{\sqrt{1-\mathrm{COS}_{\mathfrak{X}(2,1)}^2}}.
\end{aligned}
\right.
\end{equation}
\end{lemma}
\begin{proof}See Appendix~\ref{NODA_proof_reconstruction_phase_vector_fram_X}.
\end{proof}
%We set
%$$
%\bm\psi_{1(1)}=0\Rightarrow
%\cos\theta_{k(m)}=Cos\mathfrak{D}\theta_{(1,3(k-1)+m)}
%$$
%\begin{equation}
%\label{NODA_eq_dipole}
%\begin{aligned}
%&\mathcal{EQ}_{\mu}:
%\theta^*\ \mathrm{s.t.}\left\{
%\begin{aligned}
%&\mathfrak{D}\theta_{(3(j-1)+l,3(k-1)+m)}=|\theta_{j(l)}-\theta_{k(m)}|\\
%&|\theta_{j(l)}|\leq \frac{\pi}{2}\ \mathrm{for}\ j,k\in[1,n],\ l,m\in[1,3]
%\end{aligned}
%\right.
%\end{aligned}
%\end{equation}
%本来であれば，ラグランジュの未定乗数法を使ってもnonconvexというか双線形方程式は双線形のままであり，何も簡略化されない．特にDCの場合は簡略化された後，残った多項式群は自由度的に解くことができず，その最適解を満たす解が存在しない．ACの場合は解けるけどそこで止まる．しかし，ACの平均化ではそこに隠れ制約があり，それを使うとなんと双線形をconvex setに変換できる．その結果，平均的だけどsuccessfullyに解ける．
%交流を入れると問題の次数的に全ての等式制約を満たす解がfeasibleになるだけでQCQPの解法に対して寄与はしない．いずれにせよ，rank(2)として残る．もしかしたらn=2のときに限って何かがうまく作用して常にrank(2)になるかも．でもnが3以上のときはQCQPのrank-constrained問題でしかないかも
%このformulationだとrank1制約をconvex制約に緩和できる．
%Dahrouj, H. and Yu, W., 2010. Coordinated beamforming for the multicell multi-antenna wireless system. IEEE transactions on wireless communications, 9(5), pp.1748-1759.
%Huang, Y. and Palomar, D.P., 2009. Rank-constrained separable semidefinite programming with applications to optimal beamforming. IEEE Transactions on Signal Processing, 58(2), pp.664-678.
\subsection{Continuous Power-Optimal Dipole \textcolor{black}{Calculation Framework}}
We construct \textcolor{black}{sequential convex optimization-based dipole calculation framework, power-$\underline{\mathrm{O}}$ptimal $\underline{\mathrm{D}}$ipole $\underline{\mathrm{A}}$llocation (ODA), to find} a continuous family of locally power-optimal dipole solutions. This ensures smooth transitions between neighboring optimal solutions \textcolor{black}{that are suitable for the MLPs}.
\begin{lemma}
The definition of $\mathfrak{X}$ in (\ref{definition_mathfrakX}) converts the constraints in (\ref{NODA_basic_formulation}) for $i\in[1,6]$ and $j\in[1,n-1]$:
\label{NODA_lemma_ODA_constraint}
    \begin{equation}
%\label{NODA_new_constraints}
\begin{aligned}
%\forall i\in[1,6],\ \forall j\in[1,n-1],\ 
u_{cj(i)}^{a}=\frac{\mu_0}{8\pi}\mathrm{tr}\left[\mathfrak{X}\hat{K}_j^\top \hat{K}_j{\mathcal{Q}}_{j[i]}^{\top}K_j\right]
\end{aligned}
%\right.
\end{equation}
where $K_j\in\mathbb{R}^{3\times 3n}$ and $\hat{K}_j\in\mathbb{R}^{(3n-3)\times 3n}$ are constant matrices and ${\mathcal{Q}}_{j[i]}\in\mathbb{R}^{3\times 3n}$ is the matrix that are derived by $Q_{j\leftarrow k}^a$ for $k\neq j$ (See their definitions in Appendix.).
\end{lemma}
\begin{proof}See Appendix~\ref{NODA_proof_ODA_constraint}.
\end{proof}
%$$\begin{aligned}\hat{s}_{j}&\triangleq[s_{(1:3j-3)};s_{(3j+1:3n)}],\ \hat{c}_{j}=[c_{(1:3j-3)};c_{(3j+1:3n)}]\\{\mathcal{Q}}_{j[i]}&\triangleq\left[\mathcal{Q}_{{j\leftarrow 1}[i]},\ldots,\mathcal{Q}_{{j\leftarrow (j-1)}[i]},\mathcal{Q}_{{{j\leftarrow (j+1)}[i]}},\ldots,\mathcal{Q}_{{j\leftarrow n}[i]}\right].\end{aligned}$$
We convexify the primal problem in (\ref{NODA_basic_formulation}) by the standard (Shor) semidefinite program relaxation \cite{boyd2004convex} to derive the unique $\mathfrak{X}=\mathfrak{X}_0^*$ of the power-optimal solutions as follows:
\begin{equation}
\label{NODA_opt_final_0}
%\begin{aligned}
%{\mathrm{ODA}_n^{(0)}}:\ 
\min_{\substack{\mathfrak{X}\in\mathbb{R}^{3n\times 3n}}}\ \mathrm{tr}[W\mathfrak{X}]\ \mathrm{s.t.}\ 
%\left\{\begin{aligned}&
\frac{u_{j(i)}^{a}}{\mu_0/8\pi}=\mathrm{tr}\left[
\mathfrak{X}\hat{K}_j^\top \hat{K}_j{\mathcal{Q}}_{j[i]}^{\top}K_j%\ {\mathcal{R}}_{j[i]}
%\mathfrak{X}_{[3j-2:3j,[1:3j-3,3j+1:3n]]} \hat{\mathcal{Q}}_{j[i]}^{\top}
\right]%\right|\leq \varepsilon_0
%\end{aligned}\right.
%\end{aligned}
\end{equation}
for $i_{\in[1,6]}$ and $j_{\in[1,n-1]}$. Note that we divide $u_{j}^{a}$ by $\mu_0/8\pi=5e^{-8}$ to avoid excessive trial-and-error calculations. As $\mathfrak{X}_0^*$ does not necessarily satisfy the rank conditions $\mathrm{rank}(\mathfrak{X}_0^*)\leq 2$ as mentioned in Section~\ref{NODA_minimal_representation_dipole_solution}, we reduce its rank using rank-constrained optimization \cite{sun2017rank}. We solve the sequential convex programming problem for iterations $k>0$:% as follows:
\begin{equation}
\label{NODA_opt_final}
\begin{aligned}
%\mathcal{OPT}_{\mathrm{ODA}}:\ 
&{\mathrm{ODA}_n^{(k)}}:\  \mathfrak{X}_{k}^*=\argmin_{\substack{\mathfrak{X}_k\in\mathbb{R}^{3n\times 3n}, e_{k}>0}}\mathrm{tr}[W\mathfrak{X}_k]+w_k e_k\\
&\mathrm{s.t.}\left\{
\begin{aligned}
&\frac{u_{j(i)}^{a}}{\mu_0/8\pi}=\mathrm{tr}\left[
\mathfrak{X}_{k}\hat{K}_j^\top \hat{K}_j{\mathcal{Q}}_{j[i]}^{\top}K_j%\ {\mathcal{R}}_{j[i]}
%\mathfrak{X}_{[3j-2:3j,[1:3j-3,3j+1:3n]]} \hat{\mathcal{Q}}_{j[i]}^{\top}
\right],\ \forall i_{\in[1,6]},\ \forall j_{\in[1,n-1]}\\
%&P_{{\lambda}^*}\mathfrak{X}=0\\
& V_{k-1}^{*\top} \mathfrak{X}_{k} V_{k-1}^*\preceq e_{k} I_{3n-2},\quad e_{k} \leq e_{k-1}^*,
%\varepsilon &\geq \left|\frac{8\pi}{\mu_0}u_{j}^{a}(i)+\mathrm{tr}\left[(\hat{P}_{j}^\sharp P_{j})^\top\hat{\mathcal{Q}}_{j}^{i\top}\mathfrak{X}_j\right]\right|\\
%0&\leq \mathfrak{X}_j(1,1),\mathfrak{X}_j(2,2),\mathfrak{X}_j(3,3)\\ 
%J_{dN}^*&=\frac{1}{2}\mathrm{tr}\left[\left(I_3 + (\hat{P}_{j}^\sharp P_{j})^\top\hat{P}_{j}^\sharp P_{j}\right)\mathfrak{X}_j\right]\\
%J_{dN}^*&=\frac{1}{2}\sum_j \mathrm{tr}\left[\mathfrak{X}_j\right]\\
\end{aligned}
\right.
\end{aligned}
\end{equation}
where $V_{k-1}^* \in \mathbb{R}^{3n \times (3n-2)}$ denotes the orthonormal eigenvectors of $\mathfrak{X}_{k-1}^*$ associated with its $(3n\mathrm{-}2)$ smallest eigenvalues.
%from the previous iteration. 
For matrices $\mathfrak{X}^*_k$ that have converged after $k$ iterations, Lemma~\ref{NODA_lemma_reconstruction_phase_vector_fram_X} converts $\mathfrak{X}_{k}^*$ into optimal dipole solutions $\bm{m}_N^*\in\mathbb{R}^{6N}$. Our trials declared convergence when the third-smallest eigenvalue fell below a certain threshold and stabilized.
%with the submatrix of $P_{{\lambda}^*}$,  $\hat{P}_{j*}$:
%\begin{aligned}
%P_{j*}&=P_{{\lambda}^*}\ _{(:,\ 3j-2:3j)}\in\mathb{R}^{3n\times 3}\\
%\hat{P}_{j*}=\begin{bmatrix}P_{{\lambda}^*}\ _{(:,\ 1:3j-3)},&P_{{\lambda}^*}\ _{(:,\ 3j+1:3n)}  \end{bmatrix}\in\mathbb{R}^{3n\times 3n-3}.
%\hat{\mathcal{Q}}_{j}^i&=
%\begin{bmatrix}
%\mathcal{Q}_{d_{j1}}^i&\ldots&\mathcal{Q}_{d_{j(j-1)}}^i,&\mathcal{Q}_{d_{j(j+1)}}^i&\ldots&\mathcal{Q}_{d_{jN}}^i
%\end{bmatrix}.
%\end{aligned}
%where the non-symmetric matrix $\mathcal{Q}_{j\leftarrow k[i]}\in\mathbb{R}^{3\times 3}$ that satisfies $\mathrm{vec}(\mathcal{Q}_{j\leftarrow k[i]})=Q^a_{d_{jk}}(i,:)^\top$.
%\begin{remark}
%\label{NODA_conversion_X_into_sc}
%Without any loss of generality, there exists a $\delta_a^*\in\mathbb{R}^3$ corresponding to a positive vector $A^*\in\mathbb{R}_+^3$. Thus, we define $A^*$ based on the diagnal elements of $\mathfrak{X}_j$ and the relationship of trigonometric function concludes (\ref{NODA_optimal_s_a_s_L}) for associated $\delta_j^*$. We choose an arbitrary index $j$ and derives
\subsection{Learning-based Dipole Allocation \textcolor{black}{Approximation Model}}
\label{NODA_NODA_section}
This subsection summarizes the offline training method of the learning-based optimal dipole model NODA. To mitigate the curse of dimensionality for training, we chose the base frame as the line-of-sight frame $\{\mathcal{LOS}_{j\leftarrow k}\}$ of an arbitrary $k$th follower. The formulation in (\ref{NODA_time_averaged_LOS}) reduces $[r^a_{j\leftarrow k};u^{a}_{j}]\in\mathbb{R}^9$ to $[d_{j\leftarrow k};{f^{a}_{j(1,2)}};\tau^{a}_{j}]\in\mathbb{R}^6$.  Then, the NODA model takes $9(N\mathrm{-}1)\mathrm{-}3$ dimensional agent states of the specific $j_{\in\mathcal{V}_l}$th leader agents and yields the optimal $\bm{\mathtt{x}}^*$ in (\ref{NODA_mathfrak_y_reconstructing}) as follows:
\begin{equation*}
%\begin{aligned}
%\mathfrak{X}_1^*,\ {\lambda}^*
\bm{\mathtt{x}}^*
%s_{N}^*,\ c_{N}^*
=\mathfrak{M}([{r}^{l_{j\leftarrow k}},{f}^{l_{j\leftarrow k}},{\tau}^{l_{j\leftarrow k}}],{\bm\theta}) \quad\mathrm{s.t.}\quad\text{\textcolor{black}{(\ref{NODA_opt_final_0}) and (\ref{NODA_opt_final})}},%\\
%\mathfrak{X}_j^*,\ {\lambda}^*&=\mathcal{M}_{\mathrm{KC-NODA}}(\hat{r}^{l},\hat{\dot{r}}^{l},\hat{\tau}^{l},{\theta})
%\end{aligned}
\end{equation*}
%\textcolor{black}{to derive local optimal solution of (\ref{NODA_basic_formulation})}
and Lemma~\ref{NODA_lemma_reconstruction_phase_vector_fram_X} derives local optimal dipoles $\bm{s}_{N}^*$ and $\bm{c}_{N}^*$. 
\begin{remark}
Although the NODA is controller-invariant, limiting the input norm via gain tuning or guidance reduces the sample region of ${f}$ and ${\tau}$. %, thereby addressing the curse of dimensionality. 
Moreover, the distance-dependent magnetic field property naturally restricts ${r}$ to a small region.
\end{remark}
\begin{remark}
NODA is attitude-invariant when the far-field model is used. An exact model \cite{takahashi2026certified} includes the attitude information, thereby increasing the input dimensionality; otherwise, we decouple and compensate through controller design \cite{takahashi2025noda_mmh}. 
\end{remark}
\begin{remark}
\label{NODA_No_discontinuously_switching}
Discontinuously switching $\bm{m}_{k-1}{(t)}$ into $\bm{m}_{k}{(t)}$ at time $t$ induces an impulse input that may excite disturbances in the high-frequency band%or cause chattering
. The free parameter $\psi_0$ can address this problem such that $\psi_0^* =\argmin \|\bm{m}_{k-1}{(t)}-\bm{m}_{k}{(\psi_0)}\|$. %In the next subsection, we construct continuous gains using neural techniques to alleviate the additional chattering by discontinuously switching.
\end{remark}
 \begin{algorithm}[tb!]
\begin{algorithmic}[1] 
\STATE  \textbf{Inputs: }position/command $\mathbf{r}_{N}\in\mathbb{R}^{3n},\mathbf{u}_{N}\in\mathbb{R}^{6n}$%, $j\in[1, n]$ 
\STATE  \textbf{Outputs: }
%$\mu^{\mathrm{amp}*}_{N},\theta_N^*\in\mathbb{R}^{3n}$ and 
$\mu_j^{a*}(t)\in\mathbb{R}^3$, $j\in[1, n]$
\STATE  Define the base frame $\{\mathcal{A}\}$ and derive $r_j^a$ and $u_j^a$
%\STATE 距離が直径以上とCLOSのエラー対策(NODA ganerationのコード), $ODA2$で制御力一成分だけでトルク0の場合は行列のランクは1にしないとエラーが起きる．どう判別？$eig(hatP_lambda)$の0の重解-1をその値に設定.多項式の個数から判断できる
%\STATE  Solve dual problem in (\ref{NODA_opt2_N}) for $J_{\mathrm{dual}}$
\STATE  Solve ODA$_n$ in (\ref{NODA_opt_final_0}) and~(\ref{NODA_opt_final}) for $\mathfrak{X}^*$ and extract $\bm{\mathtt{x}}$ %$\mathfrak{X}^*=\ \mathcal{OPT}_{\mathrm{ODA}}\{r_{j}^{a},u_{j}^{a},P_{{\lambda}^*}\}$% on (\ref{NODA_opt_final}) with $|\frac{8\pi}{\mu_0}u_{j[i]}^{a}-\mathrm{tr}\left[\mathfrak{X}{\mathcal{R}}_{j[i]}\right]|\leq \varepsilon_0$ for numerical safety.
%\STATE  Solve $\mu^{\mathrm{amp}*}_{N},\ \theta_N^*=\mathcal{EQ}_{\mu}\{\mathfrak{X}^*\}$.
\STATE Recover $\bm{s}_{N}^*,\ \bm{c}_{N}^*$ from $\mathfrak{X}^*$ by Lemma~\ref{NODA_lemma_reconstruction_phase_vector_fram_X}% with $\forall x,y, \mathrm{COS}_{\mathfrak{X}(x,y)}=\max(-1,\,\min(1,\,\mathrm{COS}_{\mathfrak{X}(x,y)}))$ and $\sin\bm\psi_{x}=\max(-1,\,\min(1,\,\sin\bm\psi_{x}))$ for numerical safety.
\STATE Choose $\psi_0$, e.g., Remark~\ref{NODA_No_discontinuously_switching} and derive $\bm{m}(t)$%$=\mu^{\mathrm{amp}*}_j\sin(\omega_{j}t+\bm\psi_j^*+\psi_0)$
\end{algorithmic}
\caption{%Convex optimization-based power-optimal dipole allocation for 6-DoF MTQ control of $n$-agents.
power-$\underline{\mathrm{O}}$ptimal $\underline{\mathrm{D}}$ipole $\underline{\mathrm{A}}$llocation (ODA).}
\label{NODA_alg:Convex_Optimization_based_Power_Optimal_Dipole_Allocation}\end{algorithm}
\subsection{\textcolor{black}{Comparison of ODA and Nonlinear Optimizer Baseline}}
\label{comparison_ODA_nonlinear_baseline}
\textcolor{black}{
To evaluate the performance and generality of the proposed ODA, we compare it with a general nonlinear solver for our problem~(\ref{NODA_basic_formulation}). We randomly generate $n \in [3,10]$ agents and assign feasible commands $[f_{cj};\tau_{cj}]$ based on Lemma~\ref{NODA_theorem_experimental_controller}. For all trials, we compute the mean suboptimality ratio $\gamma_W$
\begin{equation}
W\triangleq I,\quad \gamma_W \triangleq ({\|[\bm{s}_N;\bm{c}_N]\|^2/2})/{J_{\mathrm{dual}}} \geq 1.
\end{equation}
and the computation time $t_{\mathrm{c}}$ [s] where $J_{\mathrm{dual}}$ is the theoretical lower bound in~(\ref{NODA_opt2_N}). As the baseline, we use MATLAB's \texttt{fmincon} 
%with a random initial guess and 
with the interior-point algorithm. %, since it is a widely used general-purpose solver for nonlinear constrained optimization. 
}

\textcolor{black}{
For each case with $n \in [3,10]$, we generate $10^4$ random samples and solve~(\ref{NODA_basic_formulation}). Following Section~\ref{NODA_Experimental_Validation}, we define a global sampling space such that $\forall j,k$, $d_{j\leftarrow k} \in [0.05, 0.32]$, $\|f_{cj}\| \in [10^{-7}, 10^{-2}]$, and $\|\tau_{cj}\| \in [10^{-8}, 10^{-3}]$. Although the primary goal of ODA is neither better optimality than the baseline nor close agreement with the global optimum, Table~\ref{ODA_general_OPT_numerical_performance_comparisons} shows that ODA achieves smaller $\gamma_W$ and $\gamma_W$ remains mild as $n$ increases. In contrast, $t_\mathrm{c}$ is several times larger than that of the baseline. However, since locally optimal solutions by ODA vary continuously with respect to the input and can therefore be approximated by an MLP, as demonstrated in Section~\ref{NODA_Experimental_Validation}, this enables a significant reduction in computation cost.
}
\begin{table}[!tb]
\centering
\caption{\textcolor{black}{Power ratio $\gamma_W\geq 1$ $[$-$]$ and computation time comparison $t_{\mathrm{c}}$ [s] using ODA and general nonlinear optimizer.% for the optimal dipole allocation problem~(\ref{NODA_basic_formulation}).
%All computations were performed on the same computing environment described in the Introduction.
}}
%\label{NODA_fig:with_table_three_MTQ}
\renewcommand{\arraystretch}{1.3}
\label{ODA_general_OPT_numerical_performance_comparisons}
\centering
\begin{tabular}{c|c|c|c|c|c|c|c|c}
\hline 
Num. $n$&3&4&5&6&7&8&9&10\\\hline
Base $\gamma_W$& 1.20 & 1.47 & 1.63 & 1.72 & 1.85 & 1.99 & 2.13 & 2.24 \\
ODA $\gamma_W$& 1.20 & 1.46 & 1.57 & 1.60 & 1.64 & 1.69 & 1.73 & 1.75 \\
\hline
Base $t_{\mathrm{c}}$& 0.3 & 0.8 & 2.2 & 5.0 & 9.7 & 17.8 & 29.3 & 48.4 \\
ODA $t_{\mathrm{c}}$& 8.3 & 12.4 & 17.9 & 25.5 & 36.5 & 54.2 & 83.7 & 139.3 \\
\hline
\end{tabular}
\label{tab:power_time_ab}
\end{table}
\section{\textcolor{black}{Illustrative Example of NODA Application}}
\label{NODA_Experimental_Validation}
This section presents an illustrative example of using the NODA model to control formation and attitude of ten MTQs. We validated our learning-based dipole allocation algorithm and multileader-based decentralization through simulations and experiments. Our goal was to achieve a regular triangle with a distance $r_d$ and its attitudes aligned with the others. 
%Ave Ave. $\int\frac{u_c-u_{\mathrm{avg}}}{u_c}$, Inte Inte$\int\frac{u_c-\int_T u(t)}{u_c}$, 
\subsection{Two-Dimensional Micro-Gravity Experiment Setup}
\label{NODA_experimental_setup_introduction}
\begin{figure}[t]
\centering
\begin{minipage}[b]{0.47\FigWidth}
    \centering
    \subfloat[Parameters of three coils.]{\includegraphics[width=0.85\linewidth]{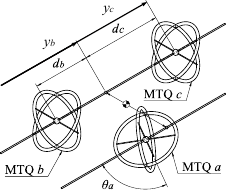}
    %\subcaption{(a) attitude}
    \label{NODA_fig:3MTQ_experiment_concept}}
\end{minipage}
\begin{minipage}[b]{0.51\FigWidth}
    \centering
    \subfloat[Experiment setup \cite{takahashi2025noda_mmh}.]{\includegraphics[width=0.85\linewidth]
    {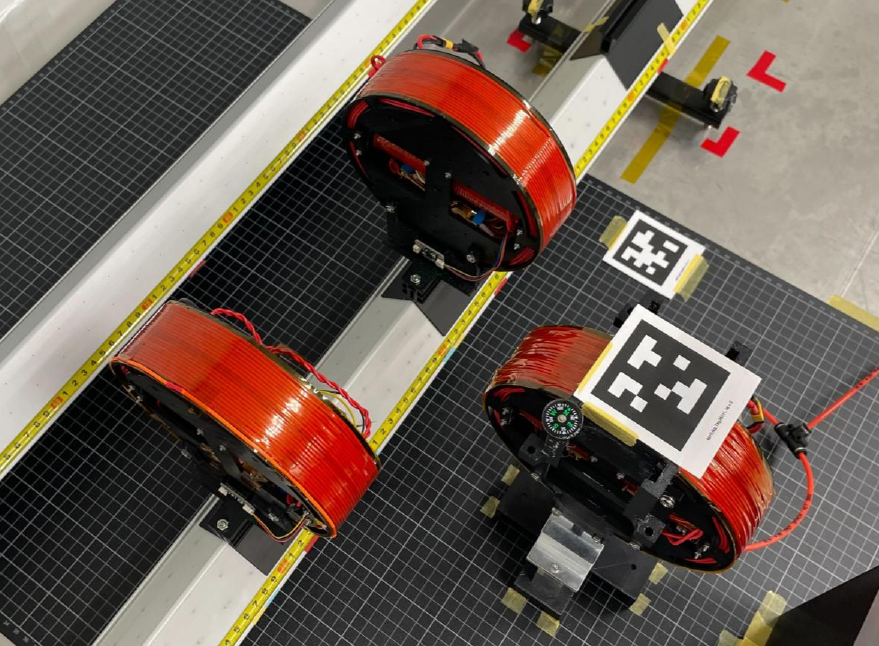}
    %{Figure/Figure_NODA/NODA3_power_experiment_output.pdf}
    %\subcaption{(a) attitude}
    \label{NODA_fig:3MTQ_setup}
    }
\end{minipage}
\begin{minipage}[b]{0.49\FigWidth}
  \subfloat[Numerical position controls.]{\includegraphics[width=1\linewidth]{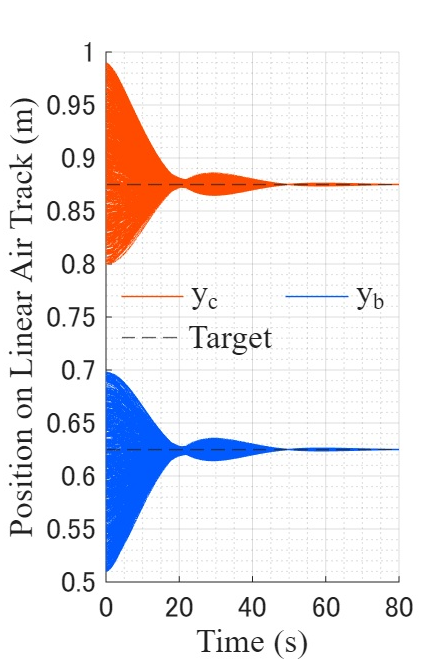}
  \label{NODA_Numerical_PDcontrol_results_positions_picture}}
\end{minipage}
\begin{minipage}[b]{0.49\FigWidth}
  \centering
  \subfloat[Decentralized position controls.]{\includegraphics[width=1\linewidth]{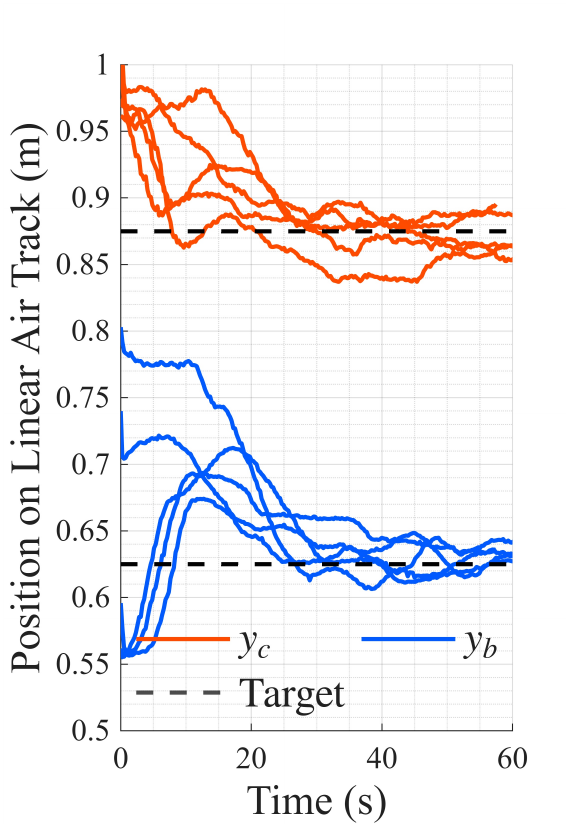}
  \label{NODA_Experimental_PDcontrol_results_positions_picture}
  }
\end{minipage}
\begin{minipage}[b]{0.49\FigWidth}
  \begin{minipage}[b]{\linewidth}
    \centering
    \subfloat[Numerical attitude control.]{\includegraphics[width=1\linewidth]{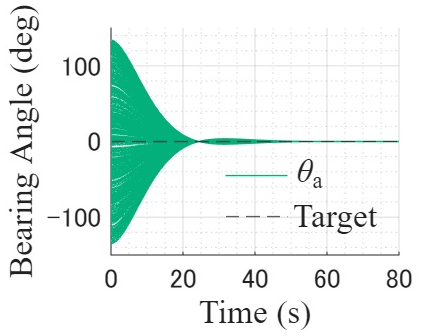}
    \label{NODA_Numerical_PDcontrol_results_attitudes_picture}}
  \end{minipage}
\end{minipage}
%\hfill
% 左半分：(a)
\begin{minipage}[b]{0.49\FigWidth}
  \centering
  \begin{minipage}[b]{\linewidth}
    \centering
    \subfloat[Decentralized attitude controls.]{\includegraphics[width=1\linewidth]{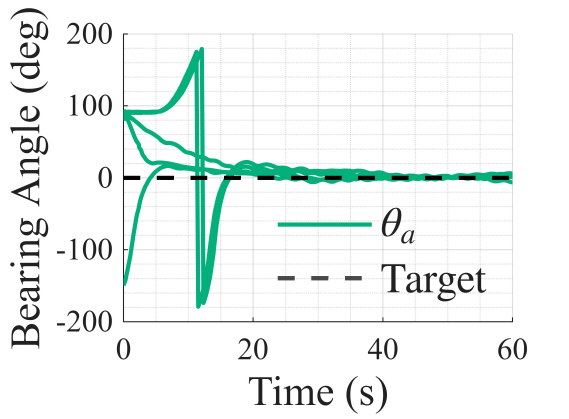}
    \label{NODA_Experimental_PDcontrol_results_attitudes_picture}
    }
  \end{minipage}
\end{minipage}
%\hfill
% 左半分：(a)
\caption{Numerical simulation results in Figs~\ref{NODA_Numerical_PDcontrol_results_attitudes_picture} and \ref{NODA_Numerical_PDcontrol_results_positions_picture} and experimental results using the linear air track and single-axis air bearing \cite{takahashi2025noda_mmh} in Figs~\ref{NODA_Experimental_PDcontrol_results_attitudes_picture} and \ref{NODA_Experimental_PDcontrol_results_positions_picture} applied the controller and decentralized allocation in subsection~\ref{NODA_controller_dipole_allocation_design}.
\label{NODA_Inverse_experiment_results}}
\end{figure}

We conduct numerical simulations replicating the ground-based environment \cite{takahashi2025noda_mmh} in Fig~\ref{NODA_fig:3MTQ_setup} as shown in Fig.~\ref{NODA_fig:3MTQ_experiment_concept}. 
%and briefly overview its setup; a full description can be found in \cite{takahashi2025noda_mmh}. 
The testbed consists of a two-axis MTQ capable of generating horizontal-plane magnetic fields with one coil incorporating an iron core. Since the microcomputers achieve time synchronization via the GPS pulse-per-second signal with 0.1 ms accuracy, we set $\omega_{j}=8\pi\times$\{1:3\} rad/s. %We describe our experimental assumptions. 
%In preparation for the 6-DoF control experiment, 
The following assumptions partially justified mechanical constraints. %of the linear track and air bearing using 
%Note that the conservation of linear momentum is rational because of the immobility of MTQ $c$. 
%\begin{assumption}The MTQs on the linear air track have virtual 3-axis RWs that cancel their generated torques, compensating for their non-rotatability. Hence, the virtual states of their angular momentum $\xi_j^{b_j}=(h_j^{b_j}-L^{b_j}/2)$ are stabilized to zero to avoid excessive bias between them.%, we assume they have the circular triaxial MTQ, and the second and third agents are equipped with 3-axis RWs. Since we should avoid the excessive bias between their angular momentum, we control the bias states about their angular momentum $\xi_j^{b_j}=(h_j^{b_j}-L^{b_j}/2)$. \end{assumption}
\begin{assumption}
\label{NODA_6DoF_control_assumption}
Forces and torques are parallel to the movable directions of the linear air track and air bearing, respectively.
%The forces act along the linear air-track direction, while torques are restricted to out-of-plane motion.
%The forces is parallel to the movable directions of the linear air track and the torques are restricted to the out-of-plane direction.
\end{assumption}
As illustrated in Fig.~\ref{NODA_fig:3MTQ_setup}, MTQs are mounted on a linear air track and a single-axis air bearing. We define the $y$-axis of the base frame along the linear track and the $z$-axis is normal to the table. %, with the positive direction upward. 
The origin was set such that the $y$ axis penetrated the geometric center of MTQs $b\&c$, and the $x$ axis completed a right-handed orthogonal frame. The positions of the MTQs were defined as $\bm{p}_{a}\triangleq[\frac{\sqrt{3}}{2}r_d;\frac{L_{\mathrm{air}}}{2};0]$ and $\bm{p}_{b,c}\triangleq[0;y_{b,c};0]$.% where $y_{b}<y_{c}$.
%, as shown in Fig.~\ref{NODA_fig:final}.
%Their desired constant values with a ``virtual RW'' were $y_{bd,cd}= (L_{\mathrm{air}}\mp r_d)/{2}$ and $\theta_{3d}=\xi_{1d}^{b_1}=\xi_{2d}^{b_2}=0$. 
%and their differential values %$\dot{y}_{1d,2d}$ and $\dot{\theta}_{3d}$ are set to 0. 
%The relative distance was measured using a time-of-flight sensor, and the AR marker information provided the angle of MTQ $a$. 
\subsection{Kinematics Controller and Dipole Allocation Design}
\label{NODA_controller_dipole_allocation_design}
We designed a formation and attitude controller using Lemma~\ref{NODA_theorem_experimental_controller} and the multileader-based control introduced in Section~\ref{NODA_NODA_subsection}. For a decentralized allocation, we separated the three coils into three groups, and each pair of MTQs, $a$--$b$, $b$--$c$, and $c$--$a$, is controlled in a decentralized manner via different frequencies. Their adjacency matrices are
\begin{equation}
%\begin{aligned}
A_1=
{\small\begin{bmatrix}
     0&1&0\\
     1&0&0\\
     0&0&0
\end{bmatrix}},A_2=
{\small\begin{bmatrix}
     0&0&0\\
     0&0&1\\
     0&1&0
\end{bmatrix}},A_3=
{\small\begin{bmatrix}
     0&0&1\\
     0&0&0\\
     1&0&0
\end{bmatrix}}
%\end{aligned}
\end{equation}
%which define centralized adjacency matrix $A=\sum_{i=1}^3A_i$. 
and their Laplacian matrix $L_i\triangleq\mathrm{Diag}(A_i\bm{1})-A_i$ with the vector of all ones $\bm{1}$ derives auxiliary commands $\mathfrak{u}_{c[i]}^a$:%=[{f}_{c[i]}^{a};{\tau}_{c[i]}^{a};{\dot{h}}_{c1}^{b_1}]\in\mathbb{R}^{24}$:
\begin{equation}
\mathfrak{u}_{c[i]}^a
=
%\begin{bmatrix}
-(L_i\otimes I_9)(K_{p}(x-x_d)+K_{d}(\dot{x}-\dot{x}_d)),
%-k_{\xi}\xi_l^{b_l}=-k_{\xi}(h_l^{b_l}- L^{b_l}_{\mathsf{g}}/2)
%\end{bmatrix}
\end{equation}
which includes $x=[p_N;\theta_N;\xi_1^{b_1};\xi_2^{b_2}]$. %\in\mathbb{R}^{24}$. 
%, position vectors $p_N$, and attitude parameters $\theta_N$. 
Lemma~\ref{NODA_theorem_experimental_controller} derives the commands for decentralized and centralized allocation: ${u}_{c[i]}^a=B_{(3,2)}^{-1}M_{(3,2)} S_{(2,2)[i]} \mathfrak{u}_{c[i]}^a$ and ${u}_c^a=\sum_{i\in\mathcal{V}_l} {u}_{c[i]}^a$ where $S_{(2,2)[i]}$ is the associated tangent space of angular momentum conservation for each pair in subsection~\ref{NODA_Kinematics_Control}.% introduced in Theorem~\ref{NODA_theorem_experimental_controller}. 

We experimentally verified this decentralized allocation for three-coil control. Each pair converts $\mathfrak{u}_{c[i]}^a$ to the dipoles based on (\ref{NODA_decentralized_allocation}) and drive the currents at $\omega_{j}=8\pi\times$\{1, 2, 3\} rad/s. The experimental result in Fig.~\ref{NODA_Inverse_experiment_results} shows that their states stabilize around the desired ones under the microgravity of the linear air track. This partially highlights the scalability extension of the multileader-based control introduced in Section~\ref{NODA_NODA_subsection}.
\begin{figure}[!tb]
\centering
% 右半分：(b) 上, (c) 下
 %\hspace*{-0.6cm} % ← 左にシフト（必要に応じて調整）
%\begin{minipage}[t]{1\linewidth}
%  \centering
\includegraphics[width=0.95\FigWidth]{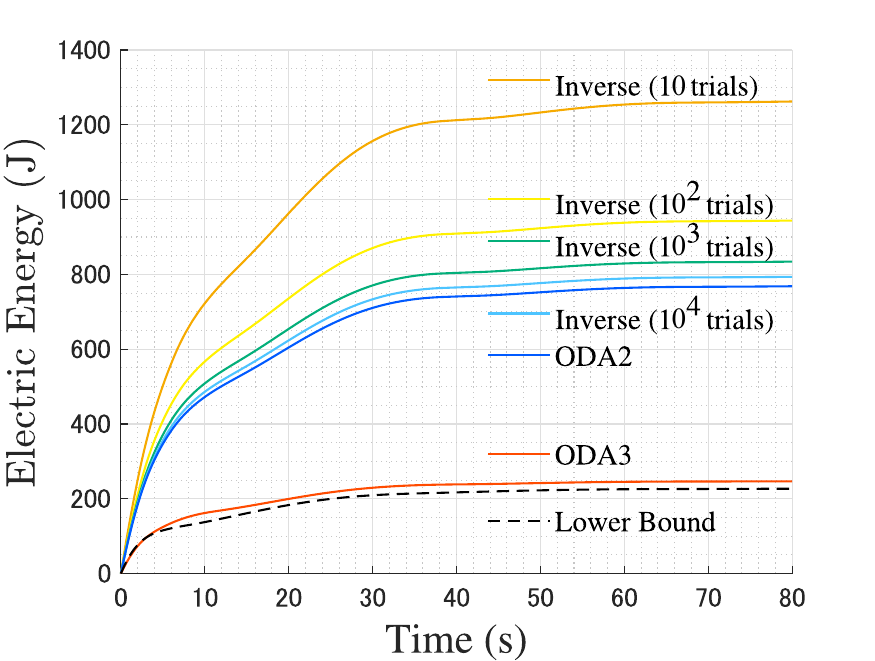}
\caption{Numerical simulation results of energy consumption of the decentralized and centralized allocations with the lower bound 
%was derived from the dual formulation of ${\mathrm{ODA}_3}$ given 
given in (\ref{NODA_opt2_N}).\label{NODA_average_power_consumption}}
  % 図には共通キャプションだけで良ければここはキャプション不要
\end{figure}
%\end{minipage}\\
%\begin{minipage}[t]{\linewidth}
%  \centering\footnotesize
%\begin{table}[tb]
%\vspace{0.5cm}
\begin{table}[!tb]
\centering
\caption{Numerical performance comparisons: power consumption and control-induced disturbance reduction.}%, and calculation time.}
\label{NODA_fig:with_table_three_MTQ}
\renewcommand{\arraystretch}{1.3}
\label{NODA_numerical_performance_comparisons}
\centering
\begin{tabular}{c|c|c|c}
\hline 
Allocations&Inverse $\times 10^4$ trials %&IR$_2$
&ODA$_2$&ODA$_3$\\%&NODA$_3$\\
%&$\times 1e^4$ trials&$\times 1e^4$&&\\\hline
\hline
Command&\multicolumn{3}{c}{$u_{c}=B_{(3,2)}^{-1}M_{(3,2)} S_{(3,2)} u_a$}\\\hline% on Theorem \ref{NODA_theorem_experimental_controller}}\\\hline
Grouping&\multicolumn{2}{c|}{Decentralized}&\multicolumn{1}{c}{Centralized}\\\hline
%Ave. [\%]&${0.092}$&-&0.095&0.69\\\hline
%Intef [\%]&${2.4}$&6.9&${2.6}$&${0.46}$\\%&${0.22}$\\\hline
Power (J)&790-1250
%&-
& 760&${250}$\\%&${220}$\\
\hline
$\Delta f$ (\%)&5.4
%&-
&5.15
&${0.40}$\\\hline%&${0.36}$\\\hline\hline
%Ave. [\%]&${0.092}$&-&0.095&0.69\\\hline%&0.6\\\hline
%Inte$\tau$ [\%]&0.9&$5.6e^3$&0.6&${1.9}$\\%&${0.70}$\\\hline
$\Delta \tau$ (\%)&5.1
%&-
&5.0&${0.53}$\\\hline%&${0.5}$\\\hline\hline
%Time [s] &${{4.2e^{-1}}/{1e^4}}$&2.5&1.0\\\hline%&${1e^{-2}}$\\\hline
%\bfseries Parameters & \bfseries Values&\bfseries Parameters & \bfseries Values\\
\end{tabular}
%\label{NODA_sec4:table1}
\end{table}
\subsection{Numerical Optimality Comparison of ODA and Baseline}
%\subsection{Controller with Centralized / Decentralized Allocation}
\label{NODA_Power_Optimal_ODA_Baseline_Comparison}
%\end{minipage}
%\caption{グラフ＋表をひとつのフロートにまとめる例}
%\end{figure}
We numerically compare the power optimality of ODA$_n$ in (\ref{NODA_opt_final_0}) and~(\ref{NODA_opt_final}) with the baseline in (\ref{NODA_decentralized_allocation}) for the three coils formation and attitude control. We conducted 500 numerical simulations using the command inputs $u_{c}$ in subsection~\ref{NODA_controller_dipole_allocation_design}. %, and magnetic field interactions were not taken into account at this stage.
Figures~\ref{NODA_Numerical_PDcontrol_results_attitudes_picture} and \ref{NODA_Numerical_PDcontrol_results_positions_picture} show the obtained time-series data, and we convert them to associated dipole via three dipole allocations: the centralized allocation ${\mathrm{ODA}_3}$ and a two decentralized allocations (inverse method in Algorithm~\ref{NODA_alg:Inverse_based_dipole_allocation} with 10$^4$ trials and ${\mathrm{ODA}_2}$). We define a power index $W_{\mathrm{power}}$:
\begin{equation}
%\label{NODA_lower_power_index}
W_{\mathrm{power}}\triangleq \int_T Rc_i^2(\tau)\frac{\mathrm{d}\tau}{T}=\frac{R}{\gamma_{\mu/c}^2}\frac{\|\bm{\overline{\mu}}_{N}\|^2}{2}
\geq \frac{R}{\gamma_{\mu/c}^2}J_{\mathrm{dual}}
\end{equation}
where $R\approx 2.1\Omega$ is the coil resistor \cite{takahashi2025noda_mmh}, $\gamma_{\mu/c}\approx 2.1$m$^2$ is the coil design ratio \cite{takahashi2025noda_mmh}, and the lower bound $J_{\mathrm{dual}}$ in (\ref{NODA_opt2_N}). Note that we memorize minimum values of the inverse method for each trials. We also derived the indexes $\Delta\eta\triangleq\Delta f$ or $\Delta \tau\in\mathbb{R}$ to investigate time-integrated control-induced disturbances:% in Section~\ref{NODA_control_induced_disturbance_reduction}:
\begin{equation}
\Delta\eta\triangleq \sup_k\sup_{j\in\mathcal{V}}\frac{\|\eta_{jc}-\sup_{t\in[t_{[k]},t_{[k]}+T]} \eta_{j(t)}\|_2}{\|\eta_{jc}\|_2}.
\end{equation}

The calculation results in Fig.~\ref{NODA_average_power_consumption} and Table~\ref{NODA_numerical_performance_comparisons} show the integration of $W_{\mathrm{power}}$ and $\Delta \{f,\tau\}$. The inverse approach showed large variations in energy consumption across $10^4$ trials and converged to a performance close to that of the ${\mathrm{ODA}_2}$ result. Notably, the centralized ${\mathrm{ODA}_3}$ achieved significantly better power efficiency, requiring only approximately one-third to one-fifth of the power consumption compared to ODA$_2$ and inverse allocation with 10 trials. 
%Notably, the centralized ${\mathrm{ODA}_3}$ achieved significantly better power efficiency than the decentralized ${\mathrm{ODA}_2}$. 
This arises from magnetic field anisotropy, implying that certain directions are more effective for generating control inputs within a given power limit. In the two-dimensional setup, the three satellites cooperatively compensate for this anisotropy, thereby enabling the derivation of a more power-optimal current solution. We also confirm the reduction of $\Delta \{f,\tau\}$ by the power optimization in Table~\ref{NODA_numerical_performance_comparisons} as proved in Subsection~\ref{NODA_control_induced_disturbance_reduction}, which reduces steady-state errors. % in both position and attitude.
\subsection{Offline Supervised Learning for NODA Construction}
\label{NODA_Learning_based_NODA}
\begin{figure}[!t]
\centering
\begin{minipage}[b]{0.493\FigWidth}
  \centering
    \subfloat[Test loss using our ODA solver.]
    {\includegraphics[width=0.9\linewidth]{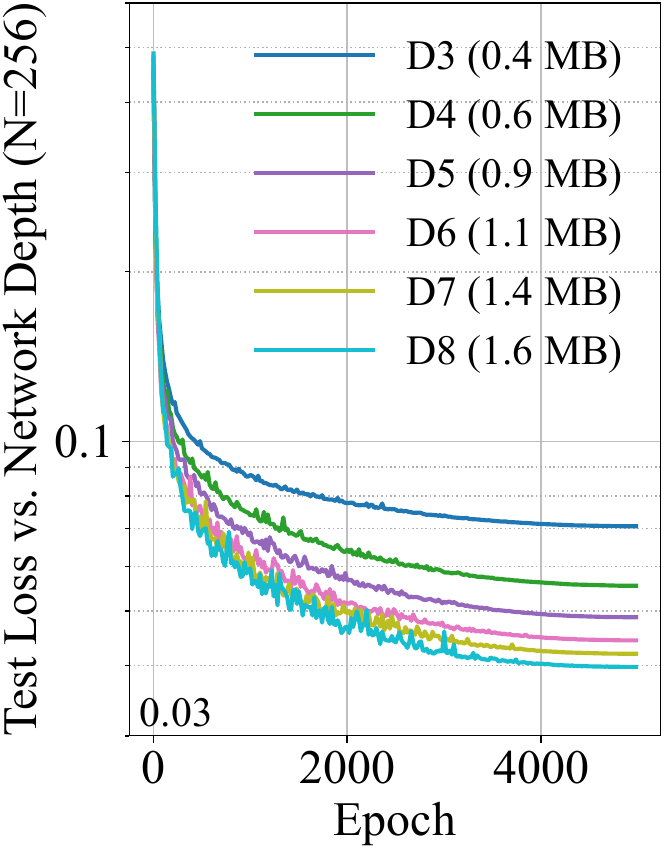}
    \label{NODA_depth_vs_loss_N256}
    }
\end{minipage}
\begin{minipage}[b]{0.493\FigWidth}
  \centering
  \subfloat[Training using general solver.\label{train_data}]{\includegraphics[width=0.9\linewidth]{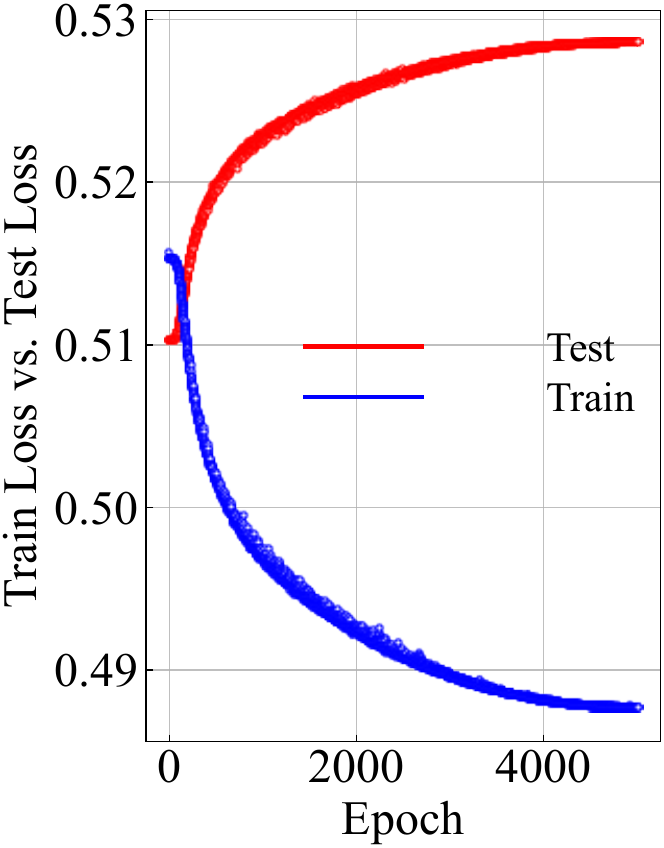}
  %depth_vs_loss_N512.}
  %\label{NODA_fig:position}
  }
\end{minipage}
\caption{\textcolor{black}{Supervised learning results using data derived by ODA and general solver in subsection~\ref{NODA_Learning_based_NODA}: (a) size–performance trade-off for ODA data, (b) train and test loss using nonlinear solver data and the best model from (a).}\label{NODA_fig:tradeoff_study}}%
\end{figure}

% using the ground-based testbed introduced in Section~\ref{NODA_experimental_setup_introduction}. 
We first train NODA $\mathfrak{M}$ for a three-coil formation and attitude control based on Section \ref{NODA_convex_allocation}. This study collected 3.5 million training samples using Algorithm \ref{NODA_alg:Convex_Optimization_based_Power_Optimal_Dipole_Allocation} for the regions shown in Fig.~\ref{NODA_Inverse_experiment_results}. We reduced the input data to the six variables:
$\chi=[y_1;y_2;{f}_{1y}^{a};{f}_{2y}^{a};{\tau}_{1z}^{b_1};{\tau}_{3z}^{b_3}]$ and the 15 dimensional label data $\mathtt{x}$ after removing redundant by momentum conservations and zero components by assumption~\ref{NODA_6DoF_control_assumption}. % indicates ${f}^{a}$ has only $y$ components and ${\tau}^{b}$ has only $z$ components. Subsequently, along with linear and angular momentum conservation, 
Our model is composed of an element-wise Leaky Rectified Linear Unit function and six ($\triangleq D$-1) hidden layers with 256 ($\triangleq N$) neurons and one hidden layer with 128 neurons based on the trade-off %between model size and test loss 
in Fig.~\ref{NODA_depth_vs_loss_N256}. %The activation function is an element-wise Leaky Rectified Linear Unit function.
%The parameters are optimized with the Adam algorithm~\cite{kingma2014adam} and The loss function is a smooth L1 objective $\mathcal{L}(\theta_w,\theta_b)=\frac{1}{6N_s}\sum_{j=1}^{N_s}\sum_{i=1}^{6}\rho_\delta\!\left(Y_i^{(j)}-\hat{Y}_i^{(j)}\right)$ where $\rho_\delta(\cdot)$ denotes the Huber loss. 
The batch sizes were $131{,}072$ 
%in the pre–training stage and $1{,}024$ during fine–tuning. 
and the learning rate followed a cosine–annealing schedule, decreasing from $10^{-3}$ to $10^{-6}$ during 5e$^3$ epoch.  %during pretraining and from $10^{-6}$ to $10^{-16}$ during fine–tuning. 
The training used a smooth L$1$-loss function that converged to a training loss of $2.66e^{-2}$ and a test loss of $3.59e^{-2}$. This result shows that the solutions derived in Algorithm \ref{NODA_alg:Convex_Optimization_based_Power_Optimal_Dipole_Allocation} successfully yield continuous local-optimal solutions to the non-convex problem in (\ref{NODA_basic_formulation}). Figure~\ref{train_data} shows that the nonlinear-solver solutions could not be approximated by MLP since their values are discontinuous with the inputs.
%\subsection{Multileader Decentralization}
\begin{figure}[!tb]
\centering
% 右半分：(b) 上, (c) 下%
%\hspace*{-0.6cm}\begin{minipage}[b]{0.2175\FigWidth}\centering\subfloat[Testbed.]{\includegraphics[width=1\linewidth]{Figure/Figure_NODA/test_rotation_position.jpg}\label{NODA_fig:testbed_setup}}\end{minipage}
%\hspace*{0.00125\FigWidth}\begin{minipage}[b]{0.325\FigWidth}\centering\subfloat[Initial condition.]{\includegraphics[width=1\linewidth]{Figure/Figure_NODA/experimental_picture_initial_3MTQ.pdf}\label{NODA_fig:initial}}\end{minipage}
%\hspace*{0.00125\FigWidth}\begin{minipage}[b]{0.325\FigWidth}\centering\subfloat[Target condition.]{\includegraphics[width=1\linewidth]{Figure/Figure_NODA/experimental_picture_final_3MTQ.pdf}\label{NODA_fig:final}}\end{minipage}
\begin{comment}
\begin{minipage}[b]{1\FigWidth}
  \centering
\subfloat[Ten MTQs belong to five groups with five AC frequencies.]% $\omega_{1,3,4,6,7}$.]
{\includegraphics[width=0.8\linewidth]{Figure/Figure_NODA/10MTQ_experiment_concept.pdf}
\label{NODA_10MTQ_experiment_picture}
    %\subcaption{(a) attitude}
    %\label{NODA_fig:attitude}
    }
  \end{minipage}
\end{comment}
  \begin{minipage}[b]{0.493\FigWidth}
    \centering
    \subfloat[Formation control results.]{\includegraphics[width=1\linewidth]{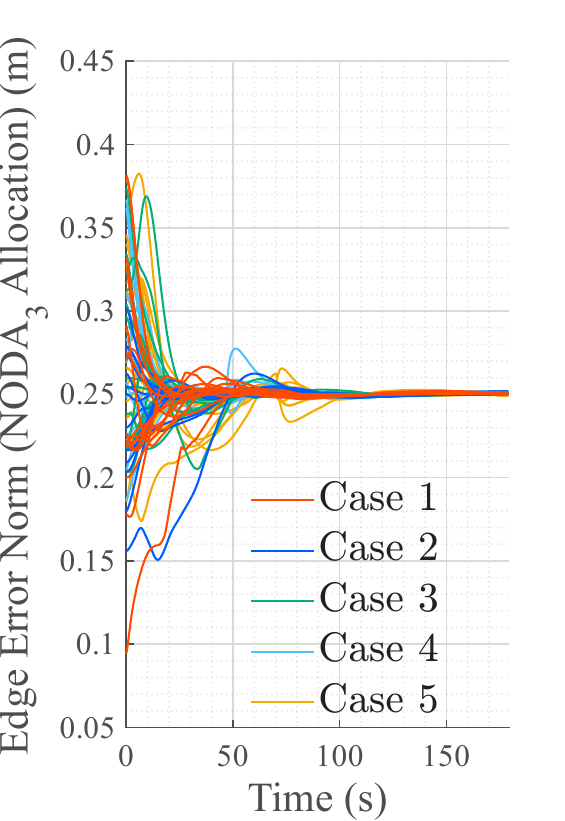}
    \label{NODA_15_edges_results}
    }
  \end{minipage}
  \begin{minipage}[b]{0.493\FigWidth}
    \centering
    \subfloat[Energy
consumption results.]{\includegraphics[width=1\linewidth]{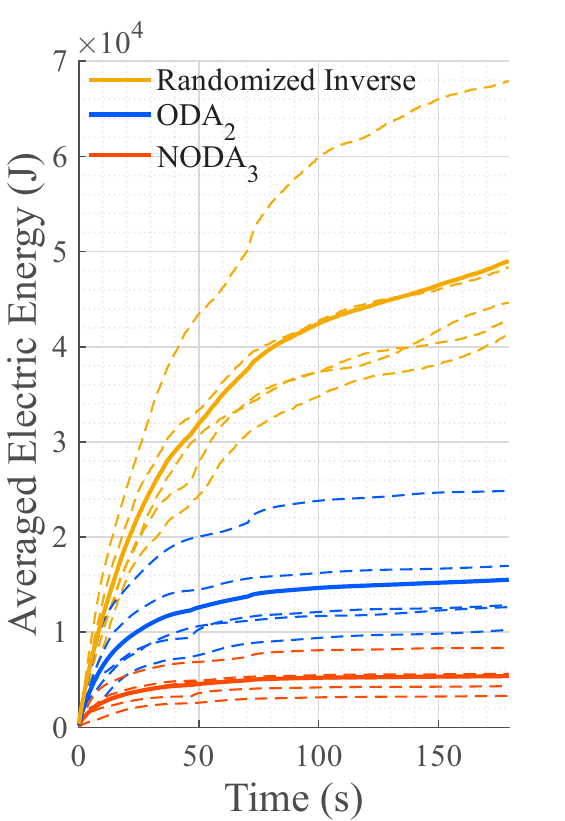}
    \label{NODA_power_consumption_NODA}
    }
  \end{minipage}
%\begin{minipage}[t]{\linewidth}\vspace{0.3cm}\captionof{table}{{Experimental performance comparison.}}\renewcommand{\arraystretch}{1.3}\label{NODA_comparison_Inverse_NODA3}\centering\begin{tabular}{c|c|c}\hline Allocations&Inverse $\times$ 1 trial &NODA$_3$\\\hline Power [J]&800-1250&${250}$\\\hline$\lim_{t\rightarrow \infty}\|p-p_d\|$ [mm]&5.4&${0.40}$\\\hline$\lim_{t\rightarrow \infty}|\theta-\theta_d|$ [deg]&5.1&${0.53}$\\\hline\end{tabular}\end{minipage}
%\end{figure}
%\begin{figure}[t]
%\centering
%\hspace{-0.05cm}
\begin{minipage}[b]{0.493\FigWidth}
    \centering
    \subfloat[Duality gap comparison of ODA$_3$.]{\includegraphics[width=1\linewidth]{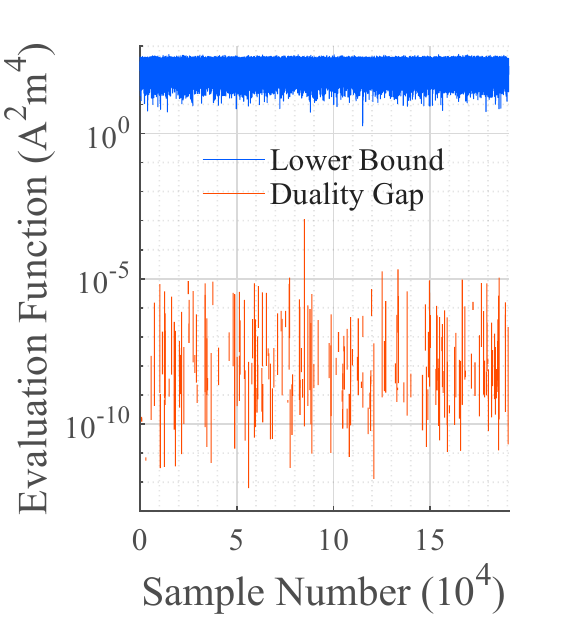}
    \label{NODA_duality_gap_picture_ODA3}
    }
  \end{minipage}
 % \hspace{0.05cm}
  \begin{minipage}[b]{0.493\FigWidth}
    \centering
    \subfloat[Allocation time comparison.]{\includegraphics[width=1\linewidth]{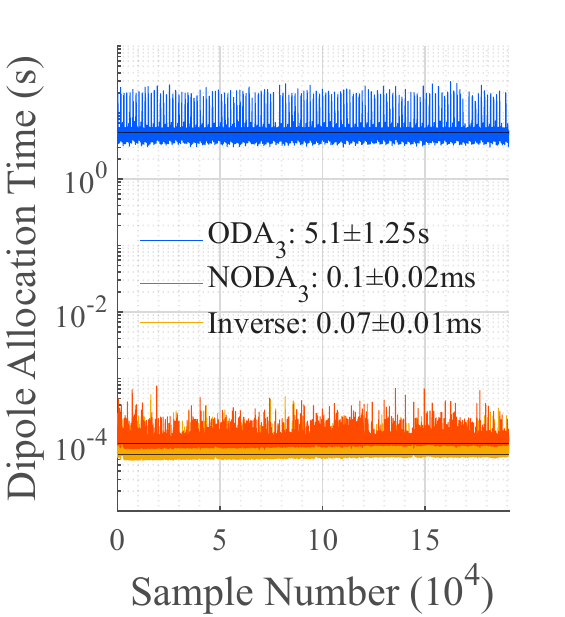}
    \label{NODA_allocation_time_picture_NODA3}
    }
  \end{minipage}
%\end{minipage}
\caption{Simulation results for multileader-based ten MTQs control using NODA with duality gap and allocation time comparison in subsection~\ref{NODA_Learning_based_NODA}.\label{NODA_NODA_experiment_results_duality_caluculation}}
%\caption{Multileader-based decentralized control overview and numerical simulation results for 10 MTQs control using trained NODA.\label{NODA_NODA_experiment_results}}
\end{figure}
\subsection{Learning-based Power-Optimal 10 MTQs Control}
\label{NODA_Learning_based_NODA}
We apply the trained NODA model to ten MTQs control and numerically demonstrate scalability via time-integrated control. We \textcolor{black}{manually} divide ten MTQs into \textcolor{black}{fixed} five groups, \textcolor{black}{each consisting of three MTQs}, managed by leaders $[1,3,4,6,7]\in\mathcal{V}_l$ using AC frequencies $\omega_{1,3,4,6,7}=8\pi\times$\{1:5\} rad/s as shown in Fig.~\ref{NODA_model_overview}. Each leader converts local state and control vectors to the power-optimal dipoles by the trained $\mathfrak{M}$ and Lemma~\ref{NODA_lemma_reconstruction_phase_vector_fram_X}. For five cases using different initial conditions, their formation and attitude converge to the desired states defined in subsection~\ref{NODA_controller_dipole_allocation_design}. We plot the control results of a total of 15 edges, the norm of the relative position, with the same case indicated by the same color in Fig.~\ref{NODA_15_edges_results}.
The power consumption was dramatically reduced by one-third to one-ten for both the initial damping phase and steady-state 
%gravity compensation 
control, as reported in Fig.~\ref{NODA_power_consumption_NODA}. These results are consistent with those reported in subsection~\ref{NODA_Power_Optimal_ODA_Baseline_Comparison}. The figures~\ref{NODA_NODA_experiment_results_duality_caluculation} present the duality gap \cite{boyd2004convex} between ODA$_3$ and its dual problem in (\ref{NODA_opt2_N}) and the allocation time comparison obtained from time-series data in the 10-agent control.  As shown in Fig.~\ref{NODA_duality_gap_picture_ODA3}, the duality gap is nearly zero, and ODA$_3$ does not lose global optimality in this planar control scenario. %, although general verification is left for future work. 
As illustrated in Fig.~\ref{NODA_allocation_time_picture_NODA3}, the proposed NODA model reduces the computation time by about 1/50,000, comparable to the inverse approach. 

\textcolor{black}{As a result, NODA achieves better optimality as the nonlinear solver with about 1/1000 of the computation time, and improves power efficiency by up to 10 times over a fast inverse-based baseline. ODA generates input-continuous solutions that are suitable for MLP approximation in Fig.~\ref{NODA_fig:tradeoff_study}, and the effectiveness and its scalability are validated in subsection~\ref{comparison_ODA_nonlinear_baseline}. These results demonstrate that the proposed learning-based framework successfully achieves both optimality and computational efficiency. Combining the multileader-based dipole allocation strategy with NODA$_{n}$ achieves power-optimal control for an arbitrary number of agents.}
\section{Conclusion}
\label{NODA_Conclusion}
This paper introduces a learning-based current calculation framework for power-optimal control of magnetically actuated systems. Because power-efficient dipole computation incurs high computational costs, we present an optimal solution calculation using sequential convex optimization and approximate it with a data-driven model, NODA. To accommodate large robot groups, a systematic decoupling architecture is employed to partition the agents into smaller units. Experimental and Numerical results show the effectiveness of our decentralization strategy, and the trained model accurately reproduces the targeted magnetic interactions while reducing power consumption, control-induced disturbances, and computation time. The proposed framework offers a practical solution for generating current commands in resource-constrained robotic systems.

%% file: appendix_NODA.tex
%\begin{comment}
\subsection{Necessary Conditions for Baseline Allocation}%in (\ref{NODA_decentralized_allocation})]
\label{Necessary_Conditions_of_Baseline}
We note the necessary conditions for the existence of the inverse in (\ref{NODA_decentralized_allocation}). Without loss of generality, we set $\theta_{F(1)}^{los}=0$ and this derives $s_{F0} = [\overline{\mu}_{F(1)};\overline{\mu}_{F(2)}\cos\theta_{F(2)};\overline{\mu}_{F(3)}\cos\theta_{F(3)}]$ and $c_{F0}=[0;\overline{\mu}_{F(2)}\sin\theta_{F(2)};\overline{\mu}_{F(3)}\sin \theta_{F(3)}]$ where $\{los\}$ are dropped in the remainder. A symbolic calculation verifies that $\overline{\mu}_{F(1)}\overline{\mu}_{F(2)}\overline{\mu}_{F(3)}\sin(\theta_{F(2)}\mathrm{-}\theta_{F(3)})\neq 0$ guarantees the existence of an inverse. 
%\begin{equation}
%\label{NODA_time-varying-dipole}
%\begin{aligned}
%{\mu}_{j}(t) ={\mu}_{\mathrm{DC}j}+{s}_j \sin (\omega_{j} t)+{c}_j \cos (\omega_{j} t)
%{\mu}_{j}(t)&=\overline{\mu}_j \sin (\omega_{j} t+\theta_j)\\&={s}_j \sin (\omega_{j} t)+{c}_j \cos (\omega_{j} t)    
%\end{aligned}
%\end{equation}
%The perpendicular vector $\mu_{F0\bot}$ of $\mu_F(t)$ based on (\ref{NODA_time-varying-dipole}) %is 
Note that dipoles with components along only two axes, i.e., whose orthogonal vectors $\mu_{F\bot}\in\mathbb{R}^3$ satisfies $\mu_{F\bot}=\delta_{il}$, which is the Kronecker delta, do not satisfy the conditions since $\mu_{F0\bot}=[\overline{\mu}_{F(2)} \overline{\mu}_{F(3)} \sin(\theta_{F(2)} - \theta_{F(3)});
\overline{\mu}_{F(3)} \overline{\mu}_{F(1)}\sin(\theta_{F(3)} - \theta_{F(1)});
\overline{\mu}_{F(1)} \overline{\mu}_{F(2)}\sin(\theta_{F(1)} - \theta_{F(2)})]$ and this leads to $\overline{\mu}_{F(l)}=0$.
%\end{comment}
\subsection{Control-Induced Disturbance Reduction via Optimization}
\label{NODA_control_induced_disturbance_reduction}
%\subsection{Proof of Lemma~\ref{NODA_quadratic_upper_bound}}
\begin{lemma}
\label{NODA_quadratic_upper_bound}% for $2\omega$ disturbance]
\textcolor{black}{
The power optimization in (\ref{NODA_basic_formulation}) reduces the upper bound of $\sup_{t\in[0,T)}\|\bm{\widetilde{d}}_{j(t)}\|$.}
\end{lemma}
%\begin{proof}See Appendix \ref{NODA_proof_quadratic_upper_bound}.\end{proof}
\begin{proof}
\label{NODA_proof_quadratic_upper_bound}
The total sinusoidal disturbance from neighbor $\bm{\widetilde{d}}_{j(t)}$ is
\begin{equation}
\label{NODA_sinusoidal_distrurbance_analytical}
\begin{aligned}
\bm{\widetilde{d}}_{j(t)}&=\sum_{k\in\mathcal{N}_j}\bm{\widetilde{d}}_{j\leftarrow k(t)}=
\begin{bmatrix}
\cos\left(2\omega t\right)\\
\sin\left(2\omega t\right)
\end{bmatrix}^\top 
\begin{bmatrix}
{d}_{c}\\
{d}_{s}
\end{bmatrix}\\
%\cos\left(2\omega t\right){d}_{c}+\sin\left(2\omega t\right){d}_{s}\\
\begin{bmatrix}
{d}_{c}\\
{d}_{s}
\end{bmatrix}
&\triangleq\sum_{k\in\mathcal{N}_j}\frac{\mu_0}{8\pi}(I_2\otimes Q_{j\leftarrow k})
\begin{bmatrix}
c_k^b\otimes c_j^b-s_k^b\otimes s_j^b\\
c_k^b\otimes s_j^b+s_k^b\otimes c_j^b
\end{bmatrix}
\end{aligned}
\end{equation}
where $\mathcal{N}_j= \{k\ |\ (j, k) \in \mathcal{E}\}$ is the neighbor set of the $j$th agent. We derive the time-supremum $\|\bm{\widetilde{d}}_{j}\|_2^2$ using (\ref{NODA_sinusoidal_distrurbance_analytical}) as
\begin{equation}
\begin{aligned}
&\sup_{t}\|\bm{\widetilde{d}}_{j}\|_2^2=\sup_{t}\ \begin{bmatrix}
\cos\left(2\omega t\right)\\
\sin\left(2\omega t\right)
\end{bmatrix}^\top
%\begin{bmatrix}\|x\|^2 & x^\top y\\x^\top y & \|y\|^2\end{bmatrix}
\begin{bmatrix}
\|{d}_{c}\|^2&{d}_{c}^\top {d}_{s}\\
{d}_{c}^\top {d}_{s}&\|{d}_{s}\|^2
\end{bmatrix}
\begin{bmatrix}
\cos\left(2\omega t\right)\\
\sin\left(2\omega t\right)
\end{bmatrix}\\
&\leq\left({\left\|
\begin{bmatrix}
    {d}_{c}\\
    {d}_{s}
\end{bmatrix}\right\|_2^2+\sqrt{(\|{d}_{c}\|_2^2-\|{d}_{s}\|_2^2)^2+4({d}_{c}^\top {d}_{s})^2}}\right)/{2}
%\\&\leq\left\|\begin{bmatrix}x\\y\end{bmatrix}\right\|_2^2
\end{aligned}.
%\end{equation}
\end{equation}
%where we use $\|v\|=1$.
%We first show $2\omega$-disturbance norm $\sup_{t\in[0,T)}\|{d}_{j\leftarrow k}^{2\omega}\|$ is linearly upper-bounded in $W_{\mathrm{power}}$.% for $m=[c_k^b;c_j^b;s_k^b;s_j^b]$.
We define $Z_0$ and $Z=Z^\top\succ 0$ such that $[c_k^b\otimes c_j^b-s_k^b\otimes s_j^b; c_k^b\otimes s_j^b+s_k^b\otimes c_j^b]=Z_0(m\otimes m)$ and $0 \preceq Z_0^\top (I_2\otimes Q_{j\leftarrow k})^\top (I_2\otimes Q_{j\leftarrow k})Z_0 \preceq  (Z\otimes Z)$. Then, $\lambda_{\max}(\cdot)\leq \mathrm{Trace}{(\cdot)}$ and the Kronecker identity \textcolor{black}{yields} $\sup_{t\in[0,T)}\|\bm{\widetilde{d}}_{j\leftarrow k}\|_2^2\leq\|[{d}_{c};{d}_{s}]\|_2^2\leq
%(m\otimes m)^\top (W\otimes W)(m\otimes m) = 
\left(\frac{\mu_0}{8\pi}m^\top Z m\right)^2$. This indicates $\|m\|^2$ linearly bounds $\sup_{t\in[0,T)}\|\bm{\widetilde{d}}_{j(t)}\|$. Therefore, the claim is proved.
\end{proof}
%This also leads to a reduction in the control-induced disturbance from an uncooperative neighbor using different frequencies, although we omit this proof for simplicity.
\subsection{Proof of Lemma~\ref{NODA_lemma_reconstruction_phase_vector_fram_X}}
\begin{proof}
\label{NODA_proof_reconstruction_phase_vector_fram_X}
A given matrix $\mathfrak{X}\in\mathbb{R}^{3n\times 3n}$ yields $\overline{\bm{\mu}}_{N}\in\mathbb{R}^{3n}$ and defines the phase difference matrix $\mathrm{COS}_{\mathfrak{X}}\in\mathbb{R}^{3n\times 3n}$ as
\begin{equation}
\label{NODA_mu_amp_N_COS_N}
\overline{\bm{\mu}}_{N}=\sqrt{\mathrm{Diag}(\mathfrak{X})},\quad \mathrm{COS}_{\mathfrak{X}}\triangleq\mathfrak{X} \oslash (\overline{\bm{\mu}}_{N}\overline{\bm{\mu}}^{\top}_{N}).
\end{equation}
We index the $3n$ components using a single index $x,y\in[1,3n]$ corresponding to $(j,l),(k,m)$: $\bm\psi_{x} \triangleq \bm\psi_{j(l)}$, $\bm\psi_{y} \triangleq \bm\psi_{k(m)}$, and $\mathfrak{X}_{(x,y)}\triangleq\mathfrak{X}_{(3(j-1)+l,3(k-1)+m)}$. Note that $\mathfrak{X}$ does not explicitly contain the redundant phases $\psi_0$:% but only their differences because
\begin{equation}
\label{NODA_no_phase_difference}
\forall\ x,y\in[1,3n],\quad \mathfrak{X}_{(x,y)}=\overline{\mu}_{x}\overline{\mu}_{y}
\cos(\bm\psi_{x}-\bm\psi_{y})
\end{equation}
Then, we can derive $\overline{\mu}_{N}\in\mathbb{R}^{3n}$ and define the phase-difference information matrix $\mathrm{COS}_{\mathfrak{X}}\in\mathbb{R}^{3n\times 3n}$ in (\ref{NODA_mu_amp_N_COS_N}). We set $\bm\psi_{1}\triangleq 0$ and $\mathrm{sign}(\sin\bm\psi_{2})\triangleq 1$ without loss of generality. The given $\mathrm{COS}_{\mathfrak{X}}$ satisfies $\mathrm{COS}_{\mathfrak{X}(x,y)}=\cos(\bm\psi_{x}-\bm\psi_{y})=\cos\bm\psi_{x}\cos\bm\psi_{y}+\sin\bm\psi_{x}\sin\bm\psi_{y}$ based on (\ref{NODA_no_phase_difference}). Subsequently, applying $y=1$ yields $\cos\bm\psi_{x}
%=\cos(|\bm\psi_{x}-\bm\psi_{1}|)
=\mathrm{COS}_{\mathfrak{X}(x,1)}$, and this recovers $\bm{c}_{N}$ in (\ref{NODA_reconstruction_c_s_N}). Moreover, (\ref{NODA_no_phase_difference}) yields 
$\mathrm{COS}_{\mathfrak{X}(x,2)}=\cos(\bm\psi_{x}-\bm\psi_{2})=\cos\bm\psi_{x}\cos\bm\psi_{2}+\sin\bm\psi_{x}\sin\bm\psi_{2}=\mathrm{COS}_{\mathfrak{X}(x,1)}\mathrm{COS}_{\mathfrak{X}(2,1)}+\sin\bm\psi_{x}\sqrt{1-\mathrm{COS}_{\mathfrak{X}(2,1)}^2}$. Since $\cos\bm\psi_{2}=\mathrm{COS}_{\mathfrak{X}(2,1)}$, we obtain $\sin\bm\psi_{x}$ in (\ref{NODA_reconstruction_c_s_N}) by applying $\sin\bm\psi_{2}=\sqrt{1-\cos^2\bm\psi_{2}}$. To satisfy $\mathrm{sign}(\sin\bm\psi_{2})\triangleq 1$, we define $\bm{s}_{N}$ in (\ref{NODA_reconstruction_c_s_N}).
\end{proof}
\subsection{Proof of Lemma~\ref{NODA_lemma_ODA_constraint}}
\label{NODA_proof_ODA_constraint}
\begin{proof}
We define $K_j$, $\hat{K}_j\in\mathbb{R}^{(3n-3)\times 3n}$, and ${\mathcal{Q}}_{j[i]}$ using $\mathcal{Q}_{j\leftarrow k[i]}$ that satisfies $\mathrm{vec}(\mathcal{Q}_{j\leftarrow k[i]})=Q^{\top}_{{j\leftarrow  k}(i,:)}$:
\begin{equation}
\begin{aligned}
K_j&= 
{\small\begin{bmatrix}
O_{3\times 3(j-1)}
&I_3&O_{3\times 3(n-j)}
\end{bmatrix}}\in\mathbb{R}^{3\times 3n}\\
\hat{K}_j&= 
{\small\begin{bmatrix}
I_{3(j-1)}&O_{3(j-1)\times 3}&O_{3(j-1)\times 3(n-j)}\\
O_{3(n-j)\times 3(j-1)}
&O_{3(n-j)\times 3}&I_{3(n-j)}\\
\end{bmatrix}}\\
{\mathcal{Q}}_{j[i]}&\triangleq
{\small\begin{bmatrix}
\mathcal{Q}_{{j\leftarrow 1}[i]}&\ldots&\mathcal{Q}_{{j\leftarrow n}[i]}
\end{bmatrix}}\in\mathbb{R}^{3\times 3}
%{\mathcal{R}}_{j[i]}&\triangleq\hat{K}_j^\top \hat{K}_j{\mathcal{Q}}_{j[i]}^{\top}K_j\in\mathbb{R}^{3n\times 3n}
\end{aligned}.
\end{equation}
Since we have $\sum_{k\neq j}\mathcal{Q}_{j\leftarrow k[i]}[{s_{k}^{a}},{c_{k}^{a}}]={\mathcal{Q}}_{j[i]}\hat{K}_j^\top\hat{K}_j[{s}_N^{a},{c}_N^{a}]$ and $[s_{k}^{a},c_{k}^{a}]\triangleq{K}_j[{s}_N^{a},{c}_N^{a}]$, Roth's column lemma 
%\cite{roth1934direct} 
derives the constraints for $i_{\in[1,6]}$th component of $u_j^{a}$ in (\ref{NODA_basic_formulation}):
\begin{equation}
%\label{NODA_new_constraints}
\begin{aligned}
&\frac{u_{j(i)}^{a}}{\mu_0/8\pi}
%&=\sum_{k\neq j}Q_{{j\leftarrow  k}(i,:)}^a({s_{k}^a}\otimes{s_{j}^a}+{c_{k}^a}\otimes{c_{j}^a})\\
%&=\sum_{k\neq j}\mathrm{tr}\left[\mathcal{Q}_{j\leftarrow k[i]}^\top(s_{j}^{a}{s_{k}^{a\top}}+c_{j}^{a}{c_{k}^{a\top}})\right]\\
=\mathrm{tr}\left[s_{j}^{a}\sum_{k\neq j}\left(\mathcal{Q}_{j\leftarrow k[i]}{s_{k}^{a}}\right)^\top +c_{j}^{a}\sum_{k\neq j}\left( \mathcal{Q}_{j\leftarrow k[i]}{c_{k}^{a}}\right)^\top\right]\\
&=\mathrm{tr}\left[\left(K_j s_{N}^{a}(\hat{K}_j{s}_N^{a})^\top +K_jc_{N}^{a}(\hat{K}_j{c}_N^{a})^\top \right) ({\mathcal{Q}}_{j[i]}\hat{K}_j^\top)^{\top}\right].
%&=\mathrm{tr}\left[\left(s_{N}^{a}{s}_N^{a\top}+c_{N}^{a}{c}_N^{a\top} \right) \hat{K}_j^\top \hat{K}_j{\mathcal{Q}}_{j[i]}^{\top}K_j\right]\\
%&=\mathrm{tr}\left[\mathfrak{X}\left(\hat{K}_j^\top \hat{K}_j{\mathcal{Q}}_{j[i]}^{\top}K_j\right)\right]\\
%&=\mathrm{tr}\left[\mathfrak{X}\hat{K}_j^\top \hat{K}_j{\mathcal{Q}}_{j[i]}^{\top}K_j\right]\\
%&=\mathrm{tr}\left[\mathfrak{X}_{[3j-2:3j,[1:3j-3,3j+1:3n]]} \hat{\mathcal{Q}}_{j[i]}^{\top}\right]
\end{aligned}
%\right.
\end{equation}   
Applying $\mathfrak{X}$ in Section~\ref{NODA_minimal_representation_dipole_solution} derives the results.
\end{proof}

%% file: main.bbl
% Generated by IEEEtran.bst, version: 1.14 (2015/08/26)
\begin{thebibliography}{10}
\providecommand{\url}[1]{#1}
\csname url@samestyle\endcsname
\providecommand{\newblock}{\relax}
\providecommand{\bibinfo}[2]{#2}
\providecommand{\BIBentrySTDinterwordspacing}{\spaceskip=0pt\relax}
\providecommand{\BIBentryALTinterwordstretchfactor}{4}
\providecommand{\BIBentryALTinterwordspacing}{\spaceskip=\fontdimen2\font plus
\BIBentryALTinterwordstretchfactor\fontdimen3\font minus
  \fontdimen4\font\relax}
\providecommand{\BIBforeignlanguage}[2]{{%
\expandafter\ifx\csname l@#1\endcsname\relax
\typeout{** WARNING: IEEEtran.bst: No hyphenation pattern has been}%
\typeout{** loaded for the language `#1'. Using the pattern for}%
\typeout{** the default language instead.}%
\else
\language=\csname l@#1\endcsname
\fi
#2}}
\providecommand{\BIBdecl}{\relax}
\BIBdecl

\bibitem{ze2022soft}
Q.~Ze, S.~Wu, J.~Nishikawa, J.~Dai, Y.~Sun, S.~Leanza, C.~Zemelka, L.~S.
  Novelino, G.~H. Paulino, and R.~R. Zhao, ``Soft robotic origami crawler,''
  \emph{Science Advances}, vol.~8, no.~13, p. eabm7834, 2022.

\bibitem{sakai2008design}
S.-I. Sakai, Y.~Fukushima, and H.~Saito, ``Design and on-orbit evaluation of
  magnetic attitude control system for the ``reimei'' microsatellite,'' in
  \emph{10th IEEE International Workshop on Advanced Motion Control}.\hskip 1em
  plus 0.5em minus 0.4em\relax IEEE, 2008, pp. 584--589.

\bibitem{takahashi2022kinematics}
Y.~Takahashi, H.~Sakamoto, and S.-i. Sakai, ``Kinematics control of
  electromagnetic formation flight using angular-momentum conservation
  constraint,'' \emph{Journal of Guidance, Control, and Dynamics}, vol.~45,
  no.~2, pp. 280--295, 2022.

\bibitem{takahashi2025noda_mmh}
Y.~Takahashi, A.~Ochi, Y.~Tomioka, and S.-I. Sakai, ``Noda-mmh: Certified
  learning-aided nonlinear control for magnetically-actuated swarm experiment
  toward on-orbit proof,'' in \emph{International Conference on Space
  Robotics}.\hskip 1em plus 0.5em minus 0.4em\relax IEEE, 2025.

\bibitem{ramirez2010new}
J.~L. Ramirez~Riberos, ``New decentralized algorithms for spacecraft formation
  control based on a cyclic approach,'' Ph.D. dissertation, Massachusetts
  Institute of Technology, 2010.

\bibitem{takahashi2026certified}
Y.~Takahashi, H.~Tajima, and S.-i. Sakai, ``Certified coil geometry learning
  for short-range magnetic actuation and spacecraft docking application,''
  \emph{IEEE Robotics and Automation Letters}, 2026.

\bibitem{shim2025feasibility}
S.~Shim, Y.~Takahashi, N.~Usami, M.~Kubota, and S.-i. Sakai, ``Feasibility
  study of distributed space antennas using electromagnetic formation flight,''
  in \emph{2025 IEEE Aerospace Conference}.\hskip 1em plus 0.5em minus
  0.4em\relax IEEE, 2025, pp. 1--18.

\bibitem{takahashi2025distance}
Y.~Takahashi, S.~Shim, and S.-i. Sakai, ``Distance-based relative orbital
  transition for palm-sized satellite swarm with guaranteed escape-avoidance,''
  in \emph{AIAA Scitech 2025 Forum}, 2025, p. 2068.

\bibitem{schweighart2006electromagnetic}
S.~A. Schweighart, ``Electromagnetic formation flight dipole solution
  planning,'' Ph.D. dissertation, Massachusetts Institute of Technology, 2005.

\bibitem{fabacher2017guidance}
E.~Fabacher, S.~Lizy-Destrez, D.~Alazard, F.~Ankersen, and A.~Profizi,
  ``Guidance of magnetic space tug,'' \emph{Advances in Space Research},
  vol.~60, no.~1, pp. 14--27, 2017.

\bibitem{abbott2017computing}
J.~J. Abbott, J.~B. Brink, and B.~Osting, ``Computing minimum-power dipole
  solutions for interdipole forces using nonlinear constrained optimization
  with application to electromagnetic formation flight,'' \emph{IEEE Robotics
  and Automation Letters}, vol.~2, no.~2, pp. 1008--1014, 2017.

\bibitem{abbasi2022decentralized}
Z.~Abbasi, J.~B. Hoagg, and T.~M. Seigler, ``Decentralized electromagnetic
  formation flight using alternating magnetic field forces,'' \emph{IEEE
  Transactions on Control Systems Technology}, vol.~30, no.~6, pp. 2480--2489,
  2022.

\bibitem{clark2012leader}
A.~Clark, B.~Alomair, L.~Bushnell, and R.~Poovendran, ``Leader selection in
  multi-agent systems for smooth convergence via fast mixing,'' in \emph{2012
  IEEE 51st IEEE Conference on Decision and Control}.\hskip 1em plus 0.5em
  minus 0.4em\relax IEEE, 2012, pp. 818--824.

\bibitem{takahashi2025scalable}
Y.~Takahashi and S.~Shin-Ichiro, ``Scalable satellite swarm deployment via
  distance-based orbital transition under $j_2$ perturbation,'' \emph{arXiv
  preprint}, 2025.

\bibitem{boyd2004convex}
S.~P. Boyd and L.~Vandenberghe, \emph{Convex Optimization}.\hskip 1em plus
  0.5em minus 0.4em\relax Cambridge University Press, 2004.

\bibitem{sun2017rank}
C.~Sun and R.~Dai, ``Rank-constrained optimization and its applications,''
  \emph{Automatica}, vol.~82, pp. 128--136, 2017.

\end{thebibliography}
